\documentclass[twocolumn,trackchanges]{aastex63}
\usepackage{amsmath}

\newcommand{\unit}[1]{\ \mathrm{#1}}
\newcommand{\Msun}{M_{\odot}}

\newcommand{\per}[2][1]{{#2}^{-#1}}
\newcommand{\rin}{R_{\rm in}}
\newcommand{\rc}{R_{\rm c}}
\newcommand{\rd}{R_{\rm d}}
\newcommand{\ud}{u_{\rm d}}
\newcommand{\St}{\mathrm{St}}

\submitjournal{ApJ}

\shorttitle{Anistropic Infall Disk Structures}
\shortauthors{Kuznetsova et al.}
\turnoffedits

\begin{document}

\correspondingauthor{Aleksandra Kuznetsova}
\email{akuznetsova@amnh.org}

\author[0000-0002-6946-6787]{Aleksandra Kuznetsova}
\altaffiliation{NHFP Sagan Postdoctoral Fellow}
\affiliation{American Museum of Natural History,
200 Central Park West,
New York, New York, 10024, USA}

\author[0000-0001-7258-770X]{Jaehan Bae}
\affiliation{Department of Astronomy,
University of Florida,
211 Bryant Space Science Center,
Gainesville, FL 32611-2055 USA}

\author[0000-0003-1430-8519]{Lee Hartmann}
\affiliation{Department of Astronomy,
University of Michigan,
 1085 S. University Ave.,
 Ann Arbor, MI 48109}

\author[0000-0003-0064-4060]{Mordecai-Mark Mac Low}
\affiliation{American Museum of Natural History,
200 Central Park West,
New York, New York, 10024, USA}

\title{Anisotropic Infall and Substructure formation in Embedded Disks}

\begin{abstract}

\par
   The filamentary nature of accretion streams found around embedded sources suggest that protostellar
   disks experience heterogenous infall from the star-forming environment, consistent with the accretion
   behavior onto star-forming cores in top-down star-cluster formation simulations. This may produce disk
   substructures in the form of rings, gaps, and spirals continuing to be identified by high-resolution imaging
   surveys in both embedded Class 0/I and later Class II sources. 
   We present a parameter study of anisotropic infall, informed by the properties of accretion flows onto protostellar cores in numerical simulations, and varying the relative specific angular momentum of incoming flows as well as their flow geometry. Our results show that anisotropic infall perturbs the disk and readily launches the Rossby wave instability (RWI). It forms vortices
at the inner and outer edge of the infall zone where material is deposited. These vortices drive spiral waves and angular momentum transport, with some models able to drive stresses \edit1{corresponding to 
   a viscosity 
parameter } on the order of $\alpha \sim 10^{-2}$. The resulting azimuthal shear forms robust pressure bumps that act as barriers to radial drift of dust grains, as demonstrated by post-processing calculations of drift-dominated dust evolution. We discuss how a self-consistent model of anisotropic infall can account for the formation of millimeter rings in the outer disk as well as producing compact dust disks, consistent with observations of embedded sources.

\end{abstract}


\section{Introduction} \label{sec:intro}

\par High-resolution imaging studies of protoplanetary disks (canonically Class II sources) in the continuum have found that many of the brightest sources are not well described by a smooth emission profile. For instance, the objects mapped in the Disk Substructures at High Angular Resolution Project (DSHARP) survey are rife with substructure in the form of rings, gaps, and spirals \citep{dsharp18}. Several studies have linked the location of ringed structures with that of potential protoplanets and the indirect kinematic effects of such planets \citep[e.g.][]{teague18, pinte19}, motivating the search for an assumed population of embedded planets \citep{bae18,dd20}. While direct detection of protoplanet candidates is still in its infancy, cases like the discovery and characterization of embedded protoplanet candidates in PDS 70 \citep{pds7018,pds7021}, lend credence to a causal relationship between planets and disk substructure. 
\par Studies of planet-disk interaction have initially assumed that gaps were predominantly opened by relatively high (Jupiter) mass planets \citep{malik15}, but recent work has shown that planets of even lower (Super-Earth) masses can not only efficiently dynamically perturb disks to create rings and gaps, but that a single planet can create the appearance of multiple concentric rings within the disk \citep{dong17,bae17}. However, dust evolution models have shown that ringed structures observed in the millimeter continuum can be produced by any mechanism that results in the trapping of grains in pressure bumps  -- regions in the disk where the pressure gradient is perturbed locally, eventually arresting the inward radial drift of dust \citep{pinilla12}. The concentration of dust in pressure bumps may trigger planetesimal formation; for example, through the streaming instability \citep{carrera21}, such that the coincidence of putative embedded protoplanets and disk substructures could also be explained by rings serving as sites of planet formation. 
\par Recently, the VLA/ALMA Nascent Disk and Multiplicity (VANDAM) survey targeting earlier phase (Class 0/I) protostellar disks, still embedded within their infalling envelopes, has also identified sources with continuum substructure \citep{sheehan20}. In some younger disks, like in the case of HL Tau which has prominent concentric ring structures \citep{alma15}, molecular line emission observations have associated flows of dense infalling gas with the presence of rings and gaps \citep{yen19}. 
Direct observations and characterization of infalling streams of filamentary gas originating outside of the dense protostellar core \citep{pineda20} have helped \emph{streamers} newly enter the vocabulary, although non-isotropy in the envelopes of early phase protostellar objects had been observed for some time before \citep{tobin10}. 

\par Given that embedded objects are regularly experiencing infall of nascent cloud material and that such infall is likely to be filamentary and anisotropic, infall-induced pressure perturbations could be a reliable mechanism for seeding substructure at early disk phases. Numerical simulations have already shown that the infall of material onto existing disks has important dynamical consequences. Studies have found that infalling material landing on disks could effectively launch the Rossby wave instability (RWI), forming dust-trapping vortices and spiral arms \citep{bae15}, aid in disk fragmentation by inducing
gravitational instability (GI) \citep{kratter08},  drive high rates of angular momentum transport \citep{lesur15}, and potentially power accretion outbursts at early times \citep{vorobyov10,bae14}.  

\par Numerical simulations of star cluster formation that follow star formation starting from the nascent molecular cloud have found that anisotropic infall is the norm rather than the exception \citep[e.g.][]{smith10,kuffmeier17,kuznetsova19}---a natural consequence of the inherent heterogeneity within filamentary star-forming environments. In \citet[][K20 hereafter]{kuznetsova20} we were able to statistically characterize the accretion behavior for all centers of gravitational collapse in our simulations by measuring the flux of mass and angular momentum into patches of cells surrounding sink particles.

We found that the accretion of cloud material onto these protostellar systems was not continuous, instead proceeding in several discrete episodes between periods of quiescence. For any given core, accretion events, often associated with spatially distinct reservoirs, have uncorrelated specific angular momenta, variable in direction and magnitude.

The results of these and similar numerical studies call into question the traditional analytic descriptions of disk formation and evolution through self-similar uniformly rotating collapse \citep[i.e.][]{tsc84}.

\par Analytic descriptions of protostellar infall typically model the deposition of mass and momentum onto disks by assuming parabolic trajectories of fluid parcels from an isotropic, uniformly rotating, spherical envelope \citep[see:][]{ulrich76,cm81}.

In the self-similar framework, the specific angular momentum of infalling material is monotonically increasing with time as outer layers of the cloud fall in, gradually building the disk outward.
In this paper, we present a modified infall model, informed by the results of K20, that takes into account stochastic variations in the angular momentum of material accreted into envelopes, as well as the filamentary and episodic nature of envelope infall. This anisotropic model is used here in a parameter study across infall properties, investigating the relative efficiency of transport and structure formation. We demonstrate how the evolution of the infall-induced RWI affects structure formation as a function of the infall properties, particularly examining the effects on disk transport (\textsection \ref{sec:transport}) and dust trapping (\textsection \ref{sec:trapping}). We show that our model of anisotropic infall results in robust structure formation and predicts a variety of ring and gap architectures, including a significant likelihood for compact dust disks.

\section{Model Description} \label{sec:methods}
\subsection{{\em FARGO3D}}
\par The simulations in this paper are performed on a two-dimensional cylindrical grid ($r$-$\phi$) with the \emph{FARGO3D} (GPU) code \citep{fargo16}. \emph{FARGO3D} solves the basic hydrodynamic equations with user-defined source terms, $\dot{\Sigma}_{\rm in}$ and $\mathbf{F}_{\rm in}$, to represent the injection of infalling mass and momentum:
\begin{equation}\label{eq:mass}
\frac{\partial\Sigma}{\partial t} +  \nabla \cdot 
(\Sigma \mathbf{u}) = \dot{\Sigma}_{\rm in},
\end{equation}
\begin{equation}\label{eq:j}
\Sigma \left ( 
\frac{\partial \mathbf{u} }{\partial t} + \mathbf{u} \cdot \nabla \mathbf{u} \right ) = -\nabla P - \Sigma\nabla \Phi + \nabla \cdot \Pi + \mathbf{F}_{\rm in},
\end{equation}
where $P = \Sigma c_{\rm s}^2$ is the isothermal pressure, $\Phi$ is the central star's gravitational potential,  and $\Pi$ is the viscous stress tensor.

\subsection{Disk Setup}
The initial disk for every run has a $\Sigma \propto \per{R}$ power-law surface density profile, with an exponential cutoff at the nominal disk radius, $\rd = 60 \unit{au}$, described by

\begin{equation} \label{eq:sig0}
    \Sigma(R) = \Sigma_0 \per{\left(\frac{R}{R_0}\right)} e^{-R/\rd},
\end{equation}
where the inner radius, $R_0 = 1 \unit{au}$, is set to $\Sigma_0 = 1700 \unit{g \ \per[2]{cm}}$ for an initial disk mass of $M_d = 0.07 \Msun$. 
\par The radial domain starts at $R_0$ and extends out to $R_{\rm max} = 260 \unit{au}$. As such, we do not directly study accretion onto the central star. At the inner boundary, we use the wave-killing boundary prescription \citep{wavekilling06} to avoid wave reflection, while we allow outflow from the outer boundary.
We adopt a temperature scaling consistent with heating by stellar irradiation such that $T \propto \per[1/2]{R}$, corresponding to a disk aspect ratio $H/R = 0.046 \ (R/R_0)^{1/4}$, which sets the sound speed $c_{\rm s}$ at each location in the radial grid \edit1{, assuming each azimuth is vertically isothermal.}
The isothermal equation of state neglects local heating through shocks or other internal heating processes. 

\par The hydrodynamic simulations are performed on a cylindrical grid of dimension $N_r = 2048, N_{\theta} = 1024$. For the purposes of capturing the development of local pressure gradients in a global simulation, we use a logarithmic radial grid corresponding to $\Delta R/ R = 0.0027$. In numerical studies of RWI vortices in protoplanetary disks, the radial half width of vortex structures depends on the local scale height and varies from $\Delta r_{\rm v} = 0.5$--$2 H$ \citep{ono18}. With the chosen $\Delta R/R$, there are at minimum 16 cells across the smallest vortices. 

\par We use the $\alpha$-disk prescription in \emph{FARGO3D} to set the viscosity $\nu = \alpha c_{\rm s} H$. To ensure that any viscous transport occurs as a result of infall-induced mechanical stresses, the background level of $\alpha$ is set to $\alpha= 10^{-7}$.

\subsection{Modeling Infall} \label{sec:methods-infall}
\par We calculate 
 the mass and momentum deposition of infalling material by adapting the analytic models in \citet[][UCM hereafter]{ulrich76,cm81}, which calculate the flux of material landing on the disk based on parabolic trajectories of fluid parcels. The fluid parcels are assumed to inherit the specific angular momentum of the infalling cloud $j_{c}$ which sets the maximum landing radius for material, the centrifugal radius
\begin{equation}
    \rc = \frac{j_{\rm c}^2}{GM_*} .
\end{equation}
In the UCM model, fluid parcels travel along streamlines in $(r,\theta)$ starting from a range of incidence angles $\theta_0 = [0,\pi]$ until landing on the disk surface where $\theta = \pi$. The landing radius of  a specific streamline with an incidence angle $\theta_0$ scales as $R = \rc \sin^2 \theta_0$. Fully 
spherical infall will deposit material from $R = [0,\rc]$. The instantaneous mass infall rate of the UCM models is then:
\begin{equation}
    \dot{\Sigma}_{\mathrm{UCM}}(R,t) = \frac{\dot{M}_{\rm in}}{4\pi R \rc} \left(1-\frac{R}{\rc}\right) ^{-1/2},
\end{equation}

where $\dot{M}_{\rm in}$ is the total mass infall rate of the cloud. 
\par We modify the UCM models to allow for filamentary infall by introducing the parameter $\rin$ to denote the inner boundary of the infall zone, which has width $\Delta R = \rc - \rin$. For the purpose of our parameter study, the modified equations are scaled such that the total integrated mass infall rate is always set to the total cloud infall rate $\dot{M}_{\rm in} = 1.0 \times 10^{-6} \unit{\Msun \ \per{yr}}$, regardless of how filamentary the accretion is. The modified infall rate 

\begin{equation}\label{eq:mass_in}
    \begin{split}
    \dot{\Sigma}_{\rm in}(R,t) & =  \dot{\Sigma}_{\mathrm{UCM}}(R,t) \left(1-\frac{\rin}{\rc}\right) ^{-1/2}\\
                           & =  \dot{\Sigma}_{\mathrm{UCM}}(R,t) \left(\frac{\Delta R}{\rc}\right)^{-1/2},\\
    \end{split}
\end{equation}
is the surface density source term in equation \ref{eq:mass}. 
The relative width of the annulus of deposited material is set by the fraction of solid angle subtended by the flow streamlines, $f$, such that
\begin{equation} \label{eq:f2}
    f^2 = \frac{\Delta R}{\rc}.
\end{equation}
\par

Momentum conservation when material with velocity $\mathbf{u}_{\rm in}$ impacts disk gas, with flow velocity $\mathbf{u}$ allows us to derive the force imparted by the infalling material (the shear source term in Equation~\ref{eq:j}) 
\begin{equation}\label{eq:j_in}
    \mathbf{F}_{\rm in}(R,t) = \dot{\Sigma}_{\rm in}(\mathbf{u}_{\rm in}(R,t) - \mathbf{u}(R,t)).
\end{equation}
The components of the radial and azimuthal flow velocities are given by the UCM projectile velocities:
\begin{equation}\label{eq:u_in}
\begin{split}
    u_{R,{\rm in}}(R,t)    & = -\left(\frac{GM_*}{R}\right)^{1/2} ,\\
    u_{\phi,{\rm in}}(R,t)  & =  \left(\frac{GM_*}{\rc}\right)^{1/2}.\\
\end{split}
\end{equation}

\par For each episode of infall, set to a duration of $t_{\rm ep} = 1.5 \times 10^4 \unit{yr}$, the simulation is evolved for an additional period of $2.0 \times 10^4 \unit{yr} $ without infall, in order to investigate the longevity of infall-produced structures during quiescence. The disk gains $\sim 0.015 \unit{\Msun}$ of material during the course of infall, about $20\%$ of its initial mass. 

\par The models used in this parameter study are detailed in Table \ref{tab:runs}. Their properties are set by the inner and outer boundaries of the infall zone, $R_{\rm in}$ and $\rc$, and the relative change in the specific angular momentum of the envelope,
    \begin{equation}
        \delta j = (j-j_0)/j_0.
     \end{equation}
The parameters were chosen based on the results of K20 to comprise a mix of filamentary and spherical infall scenarios and represent the probable range of values of $\delta j$.
\par  We assume that the initial disk surface density profile (Equation \ref{eq:sig0}) with radius $\rd$ was set by a prior episode of infall with specific angular momentum $j_0 = (GM_*\rd)^{1/2}$. 
Thus, in our model, the centrifugal radius of infall, $\rc$, parameterizes the relative change in specific angular momentum
\begin{equation}
    \delta j = \left(\frac{\rc}{\rd}\right)^{1/2} - 1.
\end{equation}

\begin{figure}[h!]
    \centering
    \includegraphics[width=2.25in]{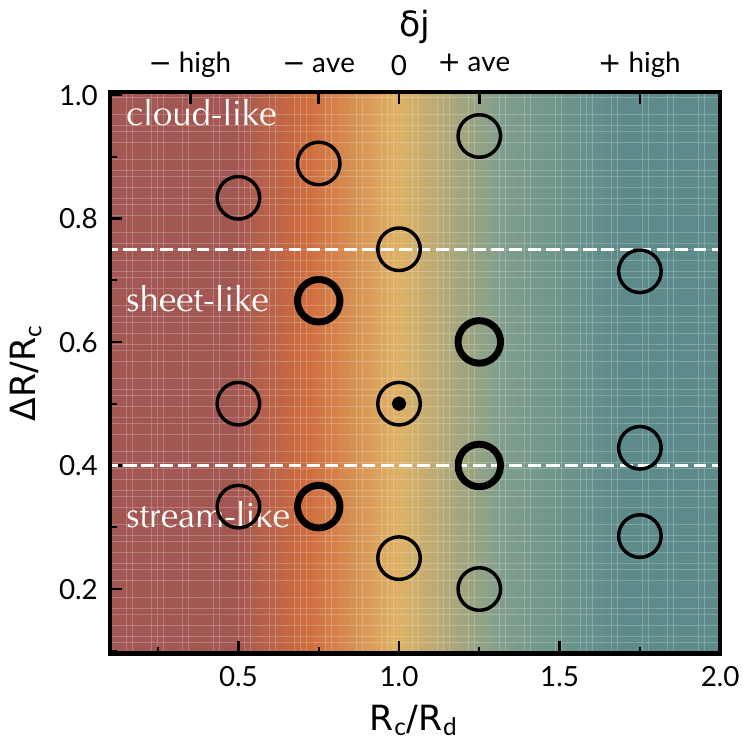}
    \caption{
The parameter space of anistropic infall explored across the variation of $\delta j$ in the infalling envelope and the filling factor of infall $f^2 = \Delta R/\rc$, where the positions of runs in the parameter space used in this paper are indicated by black edged circles. Runs corresponding to the more typical accretion events in K20 are shown as bolder circles. For reference, the fiducial run \emph{rd\_sheet} is marked with a central dot.}
    \label{fig:pspace}
\end{figure}

The results of K20 suggest that for a typical envelope accretion event of similar duration to $t_{\rm ep}$, the probable range of relative angular momentum change is $| \delta j | = 0.005 - 0.3 $. The most commonly occurring episodic accretion events have $\delta j = \pm 0.1$; these are denoted as {\em ave.} in Table~\ref{tab:runs}. The K20 runs show that 60\% of accretion events are primarily 1- or 2-dimensional, occurring in streams or sheets.   In this paper, we classify runs using the width parameter $f^2$ defined in Equation~(\ref{eq:f2}), with runs having
$f^2 \lesssim 0.4$ described as {\em stream-like}, those with \edit1{$0.4 < f^2 \lesssim 0.75$} as {\em sheet-like}, and those with $f^2 > 0.75$ as {\em cloud-like}. We choose runs from K20 that cover the relevant parameter space of anisotropic infall conditions, as shown in Figure \ref{fig:pspace}, so as to ensure each type of geometry is represented for every choice of $\delta j$. 
 
\par The infall properties, $\delta j$ and $f$, can be expressed as a function of the characteristic radii of the infall and initial disk radius: $\rc/\rd$ and $\Delta R/\rc$. In this paper, we discuss results across the parameter space in terms of whether infall occurs in the inner disk $\rc < \rd $ or outer disk $\rc > \rd$ and if accretion is stream-, sheet-, or cloud-like. 

\begin{table}[]
\centering
\begin{tabular}{lcccc}
\hline
Model         & $\rin$ {[}au{]} & $\rc$ {[}au{]} & $\delta j$ & $f$ \\ \hline
\multicolumn{5}{l}{$\delta j$ = 0}                                                              \\ \hline
rd\_stream  & 45                & 60             & 0          & stream            \\
rd\_sheet   & 30                & 60             & 0          & sheet           \\
rd\_cloud   & 15                & 60             & 0          & cloud          \\ \hline
\multicolumn{5}{l}{$-\delta j$}                                                                 \\ \hline
45\_stream  & 30                & 45             & ave.    & stream                \\
45\_sheet   & 15                & 45             & ave.    & sheet                 \\
45\_cloud   & 5                 & 45             & ave.    & cloud                  \\
30\_stream  & 20                & 30             & high       & stream                      \\
30\_sheet   & 15                & 30             & high       & sheet                 \\
30\_cloud   & 5                 & 30             & high       & cloud            \\ \hline
\multicolumn{5}{l}{$+\delta j$}                                                                 \\ \hline
75\_stream  & 60                & 75             & ave.   & stream         \\
75\_sheet   & 45                & 75             & ave.    & sheet       \\
75\_sheet   & 30                & 75             & ave.    & sheet             \\
75\_cloud   & 5                 & 75             & ave.    & cloud             \\
105\_stream & 75                & 105            & high       & stream       \\
105\_sheet  & 60                & 105            & high       & sheet         \\
105\_cloud  & 30                & 105            & high       & cloud     \\ \hline
\end{tabular}
\caption{Parameter Study Models}
\label{tab:runs}
\end{table}

\subsection{Post-processing Dust Transport}\label{sec:dust}
\par

One of our goals is to determine whether infall episodes generate ring-like substructures in the dust density by perturbing the radial pressure gradient of the disk.  In order to do this, we post-process each run at the end of the infall episode with a 1D dust transport algorithm.

\par We parameterize the pressure gradient with
\begin{equation} \label{eq:eta}
    \eta = \frac{1}{2} \left(\frac{c_{\rm s}}{v_{\rm k}} \right)^2 \frac{d\ln P}{d\ln R},
\end{equation}
where $\eta$ can be shown to be the fractional deviation of the azimuthal velocity from the Keplerian velocity $v_{\rm k}$,
\begin{equation}
   \eta = ( u_{\phi} - v_{\rm k} ) / v_{\rm k}
\end{equation}
As defined, for a typical unperturbed disk pressure gradient, $\eta$ is negative,so the gas in the disk orbits at a sub-Keplerian speed, supported by outward gas pressure.
\par Dust particles, however, do not have internal pressure, and their motions are dictated by how well-coupled they are to the gas flow. The dust-gas coupling can be described by the Stokes number $\St$, the dimensionless ratio of the drag stopping time to the local orbital time.
Particles with  $\St \ll 1$ move with the gas, while those with $\St \gg 1$  orbit closer to Keplerian speeds. For physical conditions within a typical protoplanetary disk in vertical hydrostatic equilibrium, the Stokes number of particles in the midplane as a function of the local gas surface density, \edit1{$\Sigma_{\rm g}$}, and the grain size $a$ is \citep{birnstiel16}
\begin{equation}\label{eq:Stokes}
    \St = \frac{a\rho_{\rm s}}{\Sigma_{\rm g}}\frac{\pi}{2},
\end{equation}
where $\rho_{\rm s}$ is the solid material density of the grains. 
Following \citet{w77}, the radial drift velocity of dust particles in an accretion disk depends primarily on $\eta$ and $\St$ as
\begin{equation} \label{eq:drift}
    \ud = \frac{\eta v_{\rm k}}{\St + 1/\St}.
\end{equation}

\par For simplicity, we adopt an advection dominated model of dust transport, such that the local change in the dust surface density, $\Sigma_{\rm d}$, driven by the drift velocity and gradients in the surface density and drift velocity, is
\begin{equation}\label{eq:dustad}
    \begin{split}
    \frac{\partial \Sigma_{\rm d}}{\partial t} &= - \frac{1}{R} \frac{\partial}{\partial R}\left( R \Sigma_{\rm d} \ud \right) \\
                                         &= -\Sigma_{\rm d} \frac{ \ud}{R} - \Sigma_{\rm d}\frac{ \partial{\ud}}{\partial R} - \frac{\partial \Sigma_{\rm d}}{\partial R} \ud.
    \end{split}
\end{equation}

\par In this paper, we calculate the dust transport during the post-infall period due to radial drift for 

particle sizes in sixteen logarithmically spaced bins 

in the range $\unit{[\mu m, cm]}$, at a solid density of $\rho_{\rm s} = 1.2 \unit{g \ cm ^{-3}}$. 

For the disk setup used here, all particle sizes have $\St<1$ across the disk.
We assume an initial ISM dust-to-gas ratio $ \epsilon = \Sigma_{\rm d}/\Sigma_{\rm g} = 0.01$ at $t=t_{\rm ep}$, using the azimuthal average of simulation outputs as the input, and numerically integrate the radial dust transport equation (\ref{eq:dustad}) with a first order 1D conservative upwind scheme, calculating $\ud$ at each timestep from the corresponding simulation output. These calculations are presented in \textsection \ref{sec:trapping} as a conservative estimate on the relative range and distribution of dust substructures possible as a result of this infall model. Dedicated studies of dust evolution for cases of interest will be done in future work. 

\section{Infall-induced Effects} \label{sec:results}
\subsection{Infall and the Rossby Wave Instability}
\begin{figure}[h!]
    \centering
    \includegraphics[width=0.45\textwidth]{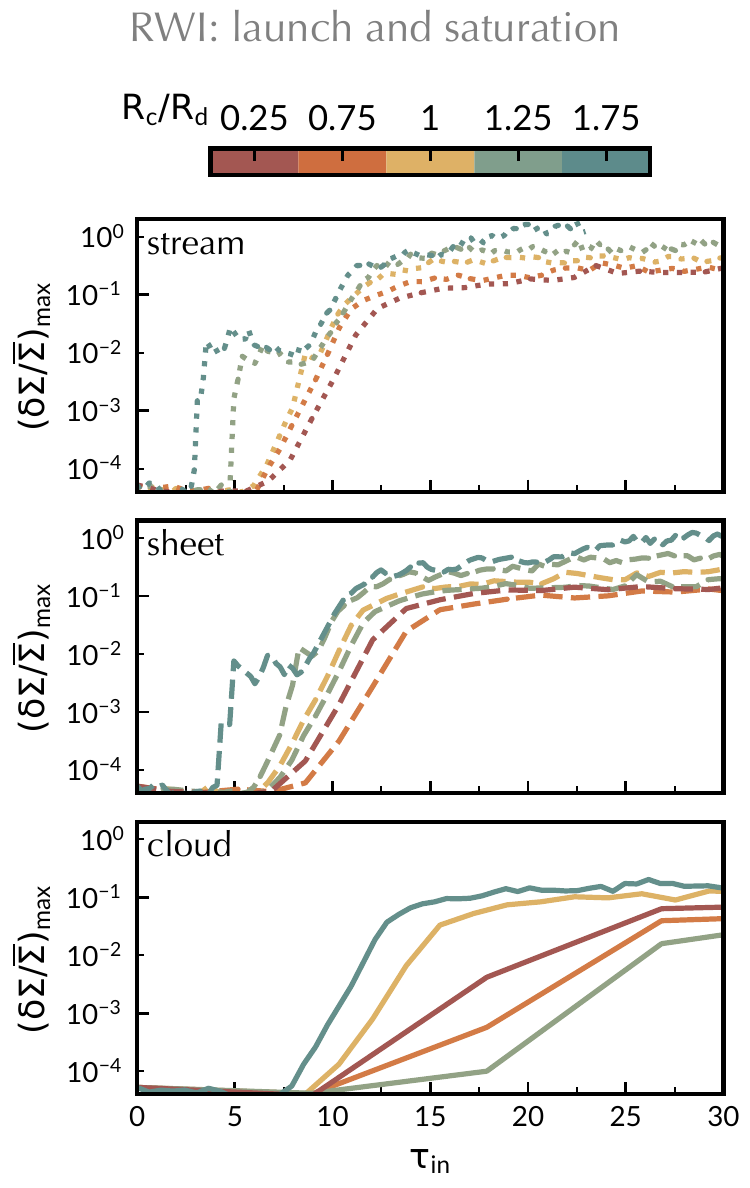}
    \caption{Growth of the maximum non-axisymmetric surface density perturbations $\delta \Sigma/\overline{\Sigma}$ scaled with the orbital timescale at the inner radius $\tau_{\rm in}$ shows the most unstable modes launch the RWI and how they proceed through the linear phase and onto non-linear growth. The color scale of the lines matches the parameters in Figure \ref{fig:pspace}, scaling with $\rc/\rd$. Similar $\Delta R/\rc$ have similar launch times \edit2{with respect to the orbital timescale at $\rin$, but a variety of growth rates} across $\rc/\rd$.}
    \label{fig:t_rwi}
\end{figure}

\par The modeled streams add mass and angular momentum within the prescribed infall zone $[\rin, \rc]$, with a radial dependence described by Equations~(\ref{eq:mass_in}) and~(\ref{eq:j_in}). While the magnitude and extent of the effect on the disk across the infall zone will vary based on the parameters of the infall, at both $\rin$ and $\rc$, where there is a transition between the unperturbed disk profile and the altered region of the infall zone, the disk will develop local bumps in the surface density and pressure gradients. These perturbations create prime conditions for the development of the RWI.

\par The traditional stability criterion of the RWI derived from linear analysis \citep{lovelace99}, requires that the disk vortensity profile
\begin{equation}
    q(R) = \frac{d/dR(R^4 \Omega^2)}{2R^3 \Sigma \Omega} 
\end{equation}
 have at least one extremum. Practically, depending on the initial disk structure, this means that local perturbations in the surface density and rotational velocity profiles both contribute to the development of the RWI. The non-linear growth and saturation of the RWI has been studied extensively with local box simulations mapping the growth rates of unstable regimes across parameter spaces of the shape, width, and amplitude of local perturbations \citep[i.e.][]{ono18}. In this study, we focus on investigating the long-term effects of the RWI on the disk structure and dynamics in the context of our prescribed physical scenario, rather than characterizing the properties of the RWI itself.

 \par In Figure \ref{fig:t_rwi}, we show the development of the phases of the RWI in terms of the growth of the fastest growing mode by plotting the maximum non-axisymmetric perturbation to the mean surface density $\delta \Sigma/\overline{\Sigma}$ over time, where $\delta \Sigma = \Sigma - \overline{\Sigma}$ and $\overline{\Sigma}$ denotes the azimuthal average of the disk surface density. The launch of the RWI occurs when the magnitude of non-axisymmetric perturbations grows past the initial noise level such that $(\delta \Sigma/ \overline{\Sigma})_{\rm max} \sim 10^{-4}$.
 
 \par During the linear regime, the growth rate---the slope of the curves in Figure \ref{fig:t_rwi}---is roughly constant. The instability saturates and becomes non-linear,resulting in the perturbations no longer growing as they approach the local disk average. Given that the global model has two regions where the RWI can develop and in some cases these regions have separations small enough for interaction, some non-linear behavior such as mode-mode coupling can be expected during what would traditionally be called the linear phase. For example, runs where the infall zone lies outside the disk radius $\rd$ show fluctuations in the growth rate. In fact, we can not expect completely analogous behavior to dedicated RWI studies, given that during infall, perturbations to the surface density and azimuthal velocity are consistently reinforced.

\begin{figure*}[ht!]
\centering
\gridline{\fig{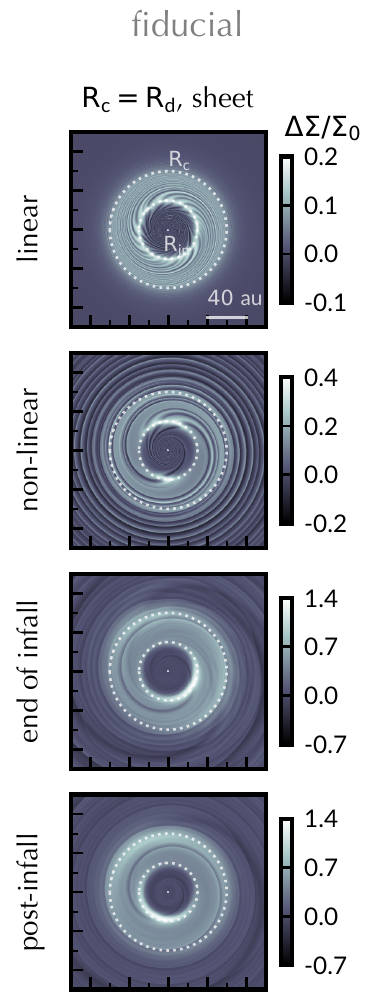}{0.25\textwidth}{}
        \fig{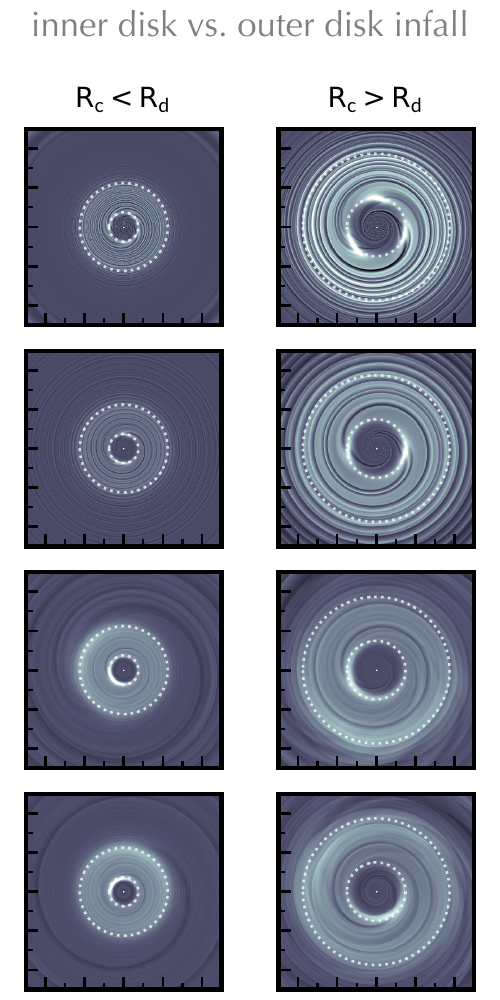}{0.33\textwidth}{}
        \fig{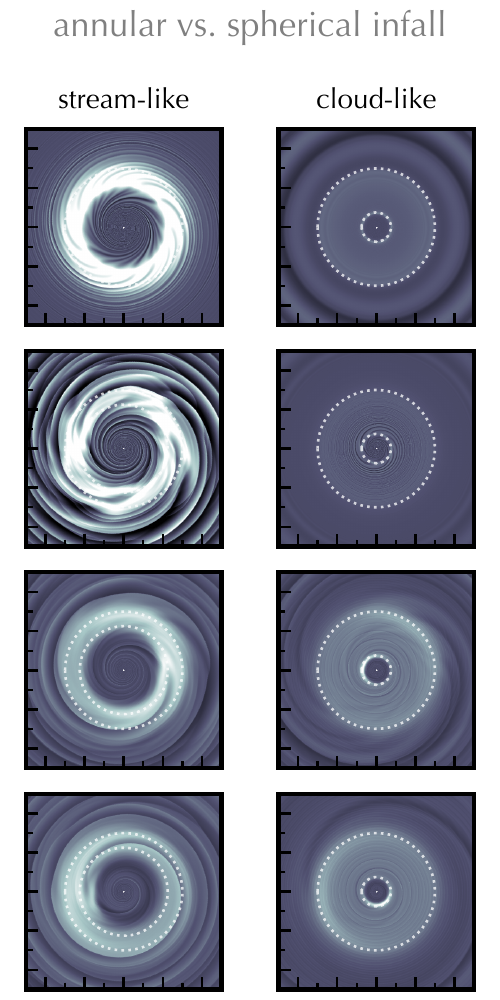}{0.33\textwidth}{}}
    \caption{The growth of the disk surface density in the central 100 au of the disk during each epoch: the linear phase of the RWI, the non-linear phase of the RWI, the end of the infall episode, and post-infall. \textit{(Left:)} The fiducial run \emph{rd\_sheet} where $\delta j = 0$ such that $\rc = \rd$ and infall is sheet-like. \textit{(Center:)} Comparison of inner and outer disk infall runs, \emph{45\_sheet} and \emph{75\_sheet}, for which $\rc < \rd$ and $\rc > \rd$ from top to bottom.  \textit{(Right:)} Comparison of runs \emph{rd\_cloud} and \emph{rd\_stream} in which the annulus width $\Delta R$ is widened or narrowed compared to the fiducial run. All panels have colorbars scaled to that of the fiducial run for each respective epoch. The boundaries of the infall zone are annotated with white dotted circles.}
   \label{fig:SD_evol}
\end{figure*}

\par While the vortex formation characteristic of the RWI proceeds more rapidly at $\rin$, the growth of the RWI is not simply dependent on the orbital timescales at $\rin$, given the presence of two zones in the disk that are liable to be unstable to the RWI. We find that the average growth rate $\gamma$ in terms of the orbital timescale at the outer radius, $\tau_{\rm c} = 2\pi \Omega^{-1}_{\rm k}(\rc)$, varies across filling factor as $\gamma \sim 0.6, 1,$ and $ 3 \ \tau_{\rm c}^{-1}$ for stream-, sheet-, and cloud-like runs respectively. Likewise, for each filling factor we find the minimum saturation time for the RWI, $t_{\rm RWI} = 8, 5,$ and $2 \ \tau_{\rm c}$.  

\par We use the nominal scaling of $t_{\rm RWI}$ and our chosen infall episode length ($t_{\rm ep} = 1.5 \times 10^4 \unit{yrs}$) to designate four main epochs of interest in our analysis and their representative times: 

\begin{itemize}
    \item the linear phase of the RWI: $t \lesssim t_{\rm RWI}$
    \item the non-linear phase: $t \sim 1.5 \ t_{\rm RWI}$
    \item the end of the infall episode: $t = t_{\rm ep}$
    \item post-infall: $t \lesssim  3.5 \times 10^4 \unit{yrs} \simeq 2.33 \ t_{\rm ep}$
\end{itemize}

\par In Figure \ref{fig:SD_evol}, we show the evolution of perturbations to the 2D surface density $\Delta \Sigma/\Sigma_0$ over the course of the four epochs. The leftmost panels show the results from the fiducial run \emph{rd\_sheet}, a sheet-like infall with $\rin = 30 \unit{au}$ and $\rc = \rd = 60 \unit{au}$. During the linear phase, perturbations are prominent within the infall zone, with vortices forming at $\rin$. As the RWI becomes non-linear, vortices at the inner edge of the infall zone merge and spiral waves are driven outside of $\rc$. At the end of infall and post-infall, $\rin$ hosts a single vortex---its spiral wake visible in contrast to the gaseous bump in the infall zone. The other panels in Figure \ref{fig:SD_evol} demonstrate how this behavior varies relative to the fiducial run if infalling material has less or more angular momentum (center) or is more or less filamentary (right). The most dramatic departures in the morphology of structures are achieved by narrowing or widening the ring. Narrow annuli of deposition, in addition to leading to a higher average surface density in the infall zone, create large-scale non-axisymmetric structures.

\begin{figure*}[ht!]
\gridline{\fig{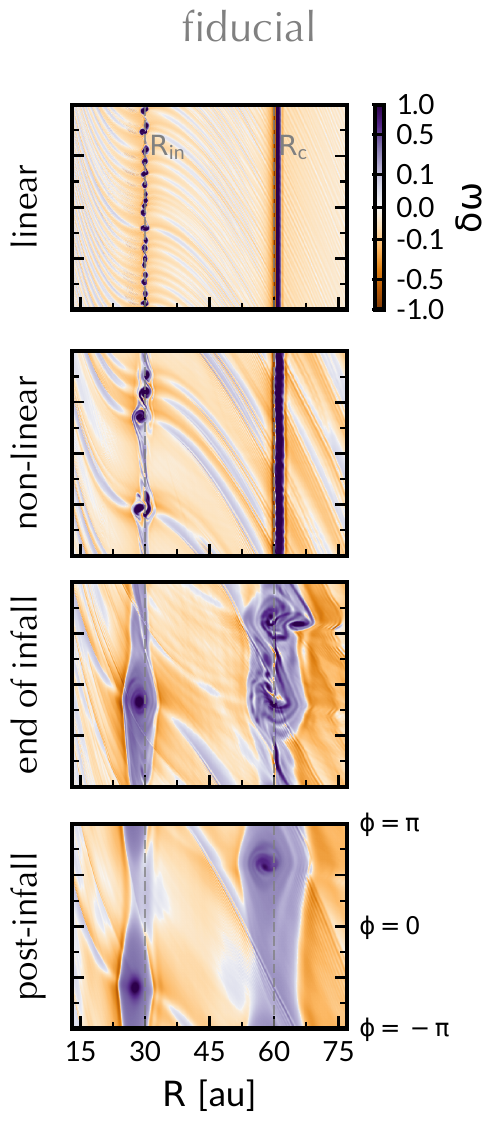}{0.25\textwidth}{}
        \fig{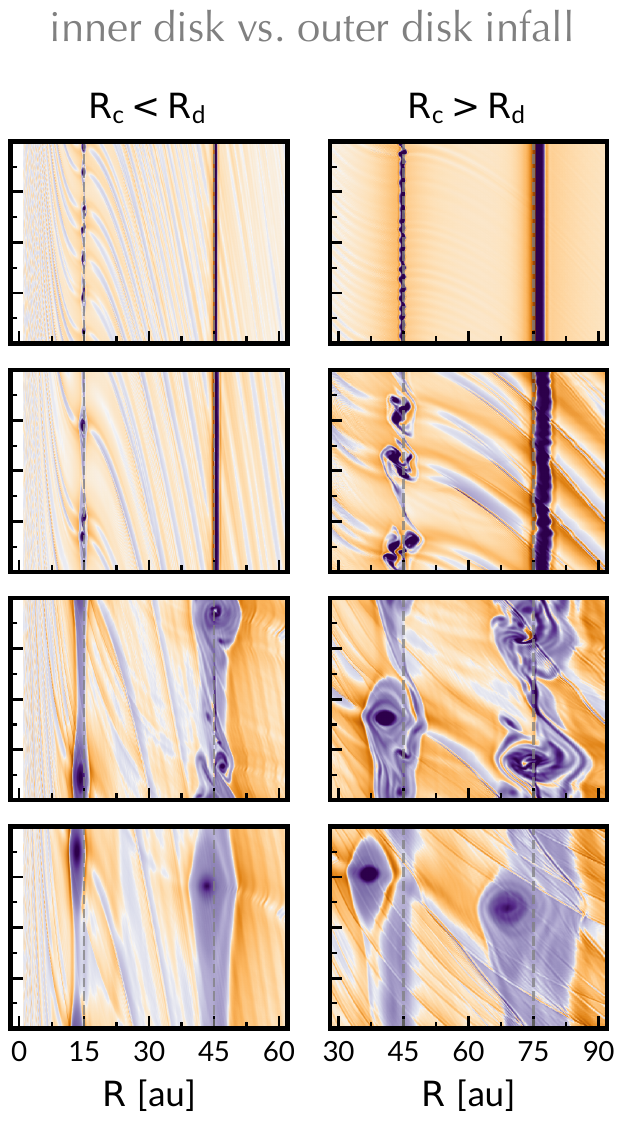}{0.32\textwidth}{}
        \fig{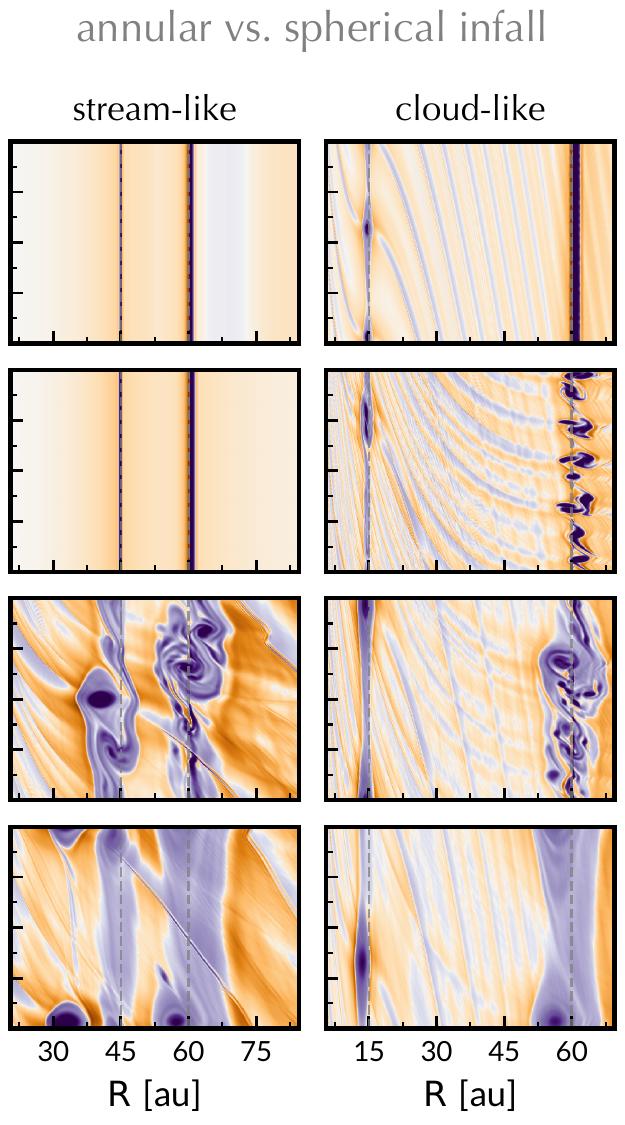}{0.32\textwidth}{}}
    \caption{Evolution of infall zone vortices and spiral features at each epoch: the linear phase of the RWI, the non-linear phase of the RWI, the end of the infall episode, and post-infall for \textit{(left)} the fiducial run \emph{rd\_sheet} such that $\rc = \rd$ and infall is sheet-like; \textit{(center)} runs \emph{45\_sheet} and \emph{75\_sheet} for inner and outer disk infall respectively; and \textit{(right)} runs \emph{rd\_cloud} and \emph{rd\_stream} in which the annulus width $\Delta R$ is widened or narrowed compared to the fiducial run. Each panel is in $\phi$-$r$ coordinates, 60 au wide, and centered on the infall zone. Vortex centers have normalized, density-weighted vorticity $\delta \omega \sim 1$, while spiral features are visible as alternating stripes of $\delta \omega \pm 0.1$.}
    \label{fig:vortices}
\end{figure*}

\par RWI vortices are visible in Figure \ref{fig:SD_evol} as localized surface density perturbations at the edges of the infall zones. 
We use the vorticity, $\omega = \nabla \times \bf{u}$, to define a normalized density-weighted vorticity factor,
\begin{equation}
    \delta \omega = \Sigma (\omega-\omega_0)/(\Sigma_0 \omega_0)
\end{equation} 

plotted in Figure \ref{fig:vortices} at the four main epochs for the same set of selected runs as in Figure \ref{fig:SD_evol} in $\phi$-$r$ space. With this normalization, the centers of vortices have $\delta \omega \sim 1$ and alternating regions of $\delta \omega \sim \pm 0.1$ highlight the peaks and troughs of spiral waves.
\par During the linear phase, individual vortices form at the edges of the infall zones, each vortex appearing to be the foot point for a spiral wave. With vortex formation occurring at both the inner and outer edges of the infall zone, spiral density waves are launched from both $\rin$ and $\rc$, resulting in interference within the infall zone itself. As the instability grows, smaller denser vortices merge into larger, more diffuse structures, with fewer spiral arms. 
Merged vortices are wider at outer radii, their radial extent scaling with the scale height H at vortex center.

\par While most vortices at the inner edge of the infall zone merge into a single vortex within the infall episode period, runs with the same $\rc$ do not necessarily share the same timescales for vortex merging. The average number of total vortices across each run depends on the orbital timescale of the midpoint between $\rin$ and $\rc$, suggesting that interactions between modes in narrower rings of infall can affect the evolution of vortex structures. Even so, interference between spiral waves within the infall zone can still be seen for the widest separations.

\par Through the mechanism of the RWI, annular infall has implications for two major disk processes: inducing turbulence and thereby transport of disk materials (\textsection \ref{sec:transport}) and the creation of pressure bumps liable to trap dust particles by arresting radial drift (\textsection \ref{sec:trapping}). We explore the longevity of structures across our parameter space after infall in \textsection \ref{sec:after_in}.

\subsection{Turbulence and Transport}\label{sec:transport}
\par
\begin{figure*}[ht!]
\centering
\gridline{\fig{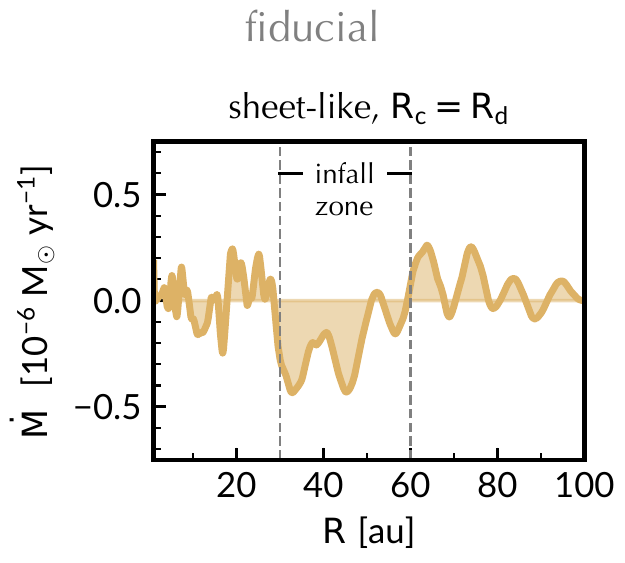}{0.22\textwidth}{}
 \fig{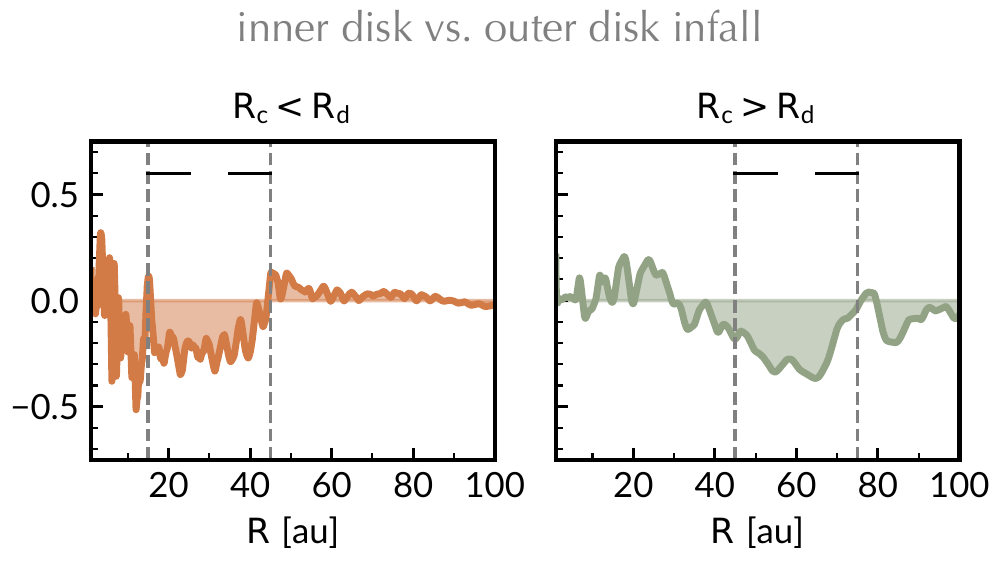}{0.35\textwidth}{}
\fig{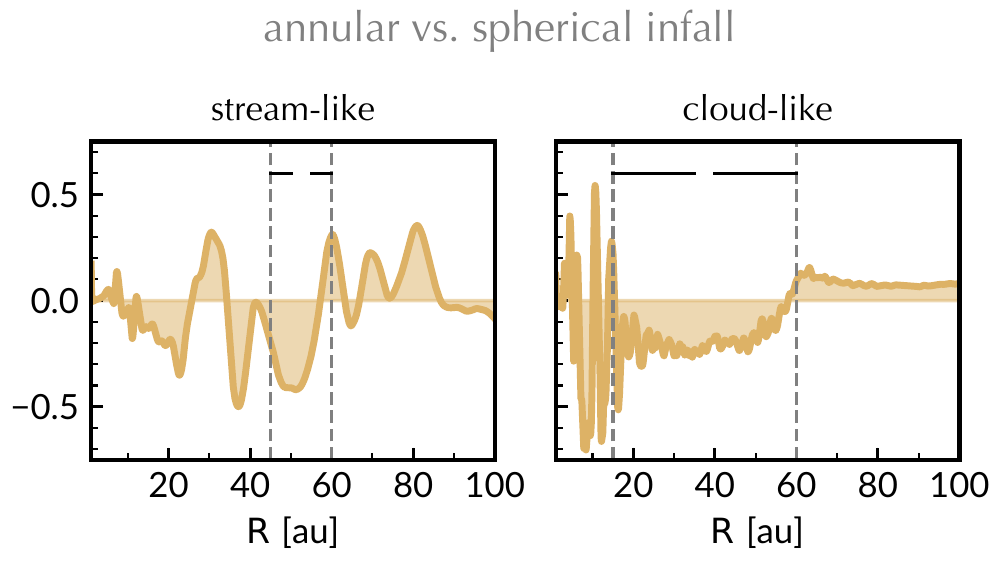}{0.35\textwidth}{}}
\caption{The average radial mass flux in units of total mass infall rate, $\dot{M}_{\rm in} = 10^{-6} \Msun \unit{yr^{-1}}$, across the inner 100 au at the end of infall---time averaged across 0.3 kyr---where negative values indicate inward mass transport. \textit{(Left:)} the fiducial run, sheet-like accretion with $\rc = \rd$. \textit{(Center:)} sheet-like runs with inner disk $\rc < \rd$  and outer disk $\rc > \rd$ infall. \textit{(Right:)} cloud-  and stream- like infall with $\rc = \rd$. }
    \label{fig:mdot}
\end{figure*}

 During infall, the bulk of mass transport is driven directly by infall as shown by the average radial mass flux within the disk:
\begin{equation}
    \dot{M}(R) = 2 \pi R \overline{\Sigma u_R}
\end{equation}
for our comparative set of models (Figure \ref{fig:mdot}). Inward mass flux---where $\dot{M}(R) < 0$ ---is at most 20--30\% of the total mass infall rate $\dot{M}_{\rm in} = 10^{-6}  \Msun \ \unit{yr}^{-1}$.

\par Exterior to $\rc$, mass flows outward, contributing to spreading of the gas ring accumulating within the infall zone. Interior to $\rin$, in the innermost regions of the disk, the radial mass flux is dominated by wave transport, particularly for models with inner disk infall such that $\rc < \rd$. We expect that spiral waves may transport angular momentum outward, thus allowing further mass accretion to occur. However, during the infall stage mass flux in these inner regions is at most 10\% of $\dot{M}_{\rm in}$. At this stage, the time average of the mass flux within the outer extent of the infall zone is between  $\dot{M}(<\rc) \sim 10^{-8}$--$10^{-7} \Msun \unit{yr}^{-1}$ for all disks,  with models of stream-like infall experiencing the most fluctuation within that range. 

\par In addition to the injection of momentum in the radial direction, the impact of infalling material onto the disk generates azimuthal shear (see Eq. \ref{eq:u_in}). Shear stresses, also called Reynolds stresses, can drive turbulence. We calculate the turbulent viscosity parameter
$\alpha$ due to Reynolds stresses at each radius:
\begin{equation}{\label{eq:alpha}}
    \alpha_{\rm Rey}(R) = \frac{\int \Sigma \delta u_R \delta u_\phi \ R \ d\phi}{\int \Sigma c_{\rm s}^2 \ R \ d\phi}
\end{equation}
where $\delta u_R$ and $\delta u_{\phi}$ are the residual velocities from the azimuthal average, defined by $\delta u = u - \overline{u}$. The disk-wide $\alpha_{\rm Rey}$ is computed as the radial average of Equation~(\ref{eq:alpha}). The total $\alpha$ -- a sum of the induced Reynolds $\alpha$ and the background viscosity parameter: $\alpha = \alpha_{\rm Rey} + 10^{-7}$ -- is an average estimate for the turbulent stresses acting over time at a particular radius in the disk; the shear stresses produced by spiral waves will vary over azimuth.

\par
\begin{figure}[h!]
    \centering
    \includegraphics[width=0.9\columnwidth]{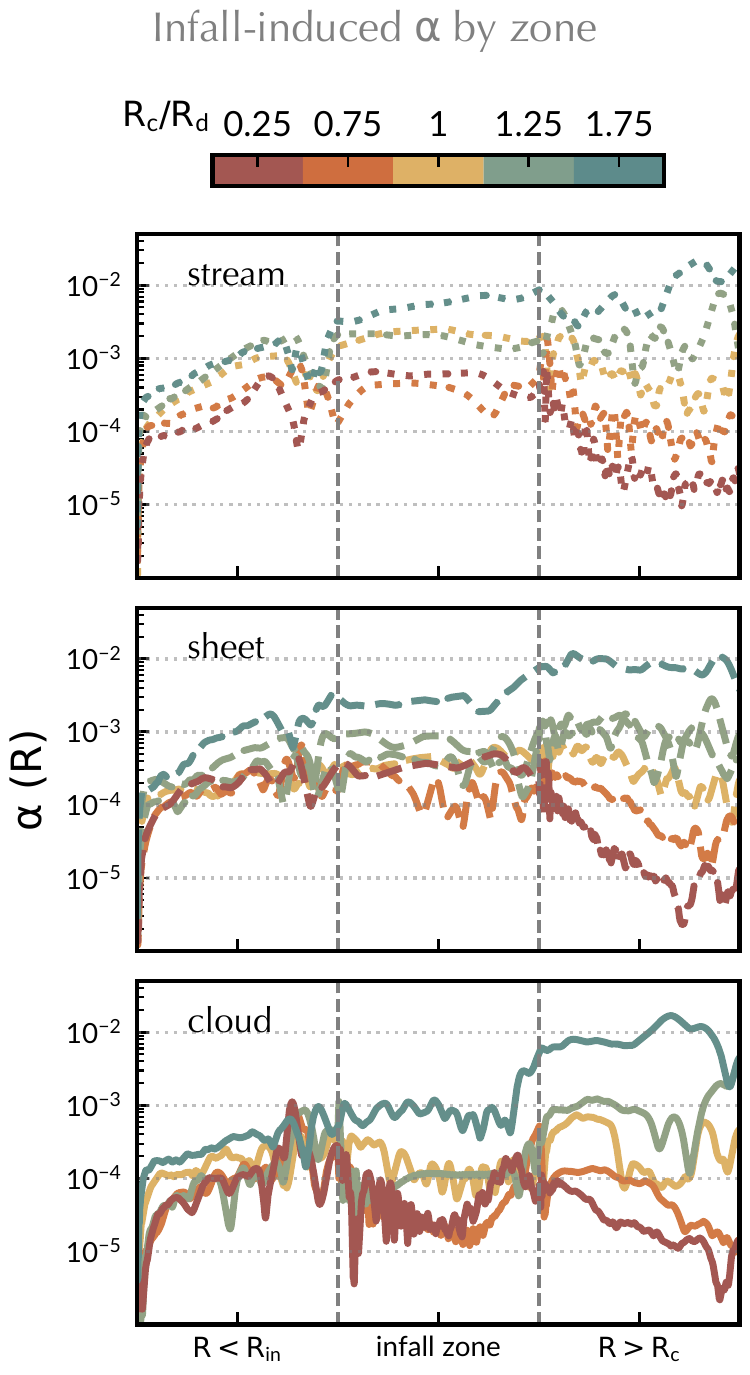}
    \caption{Measured turbulence due to infall across three zones of the disk, $R\leq \rin$, $\rin< R 
    \leq \rc$, and $R >\rc$, shown here with each zone scaled to the same apparent width for illustrative purposes. Panels from top to bottom show $\alpha$ at $t=t_{\rm ep}$ for inner disk ($\rc < \rd$), fiducial ($\rc = \rd$), and outer disk ($\rc > \rd$) infall, respectively.}
    \label{fig:alpha_iz}
\end{figure}

 In the prescribed infall model, azimuthal shear is introduced across the entire infall zone, with the largest velocity shear initially at the inner edge of the infall zone $\rin$. Once the RWI develops, spiral arms driven by vortices contribute to the total mechanical shear stress. For stream-like infall with narrow infall zones, interference between spiral waves drives turbulence within the infall zone (Figure \ref{fig:alpha_iz}), raising the overall level of $\alpha$.
\par In the case of inner disk infall where $\rc/\rd < 1$, $\alpha$ decreases outside of the infall zone, in direct contrast with outer disk infall ( $\rc/\rd > 1$), where $\alpha$ grows outward,
peaking much further out in the disk, well outside of the infall zone. The fiducial case where $\rc = \rd$ produces the most constant distribution of $\alpha$ across the disk.

\par Stream and sheet-like infall produces, on average, larger $\alpha$ values than more spherical deposition, but the location of the centrifugal radius $\rc/\rd$ plays the most important role in determining the disk-wide $\alpha$. Outer disk infall drives the largest amount of turbulence across the largest fractions of the disk, with values reaching $\alpha \sim 10^{-3}$--$10^{-2}$, around two orders of magnitude greater than the typical values with inner disk infall.

\par
\begin{figure}
    \centering
    \includegraphics[width=0.9\columnwidth]{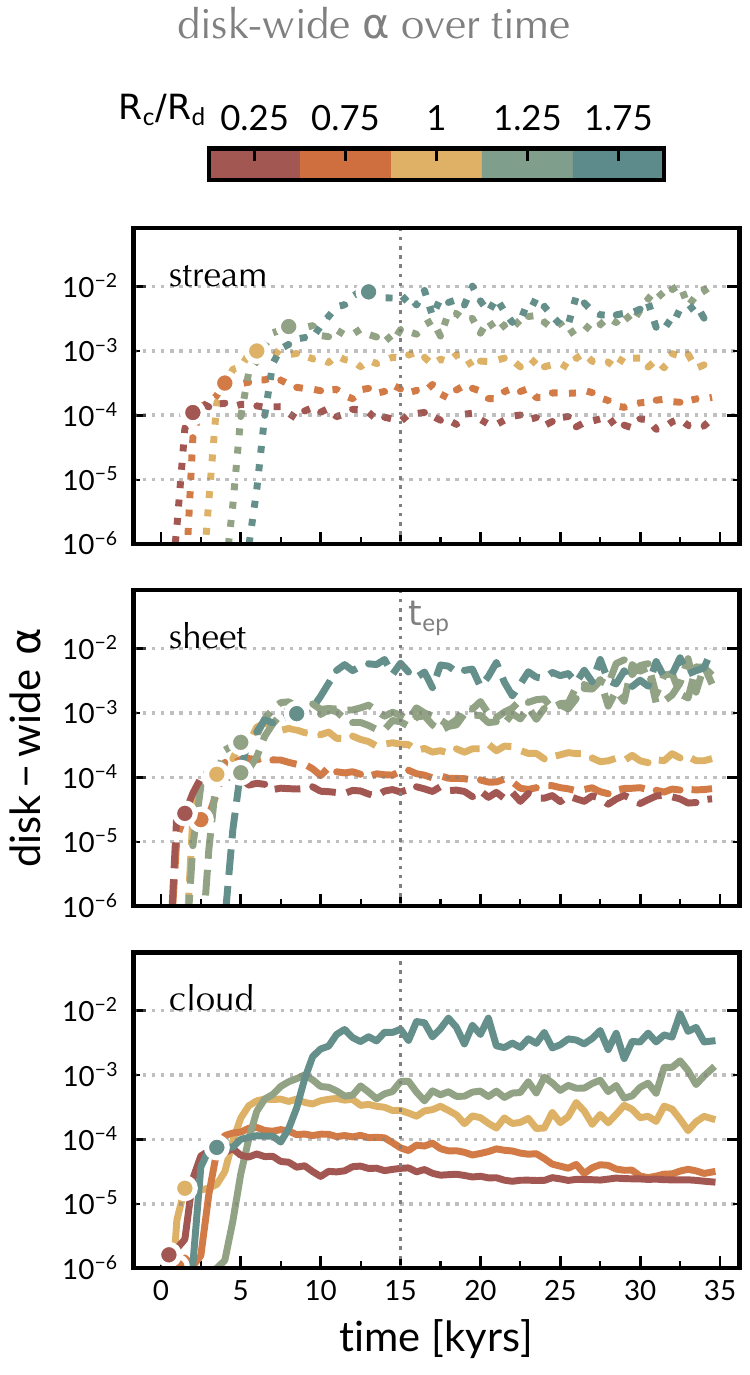}
    \caption{ The disk-wide measure of the turbulent $\alpha$ parameter, over the course of simulation runtime. Circles mark $t_{\rm RWI}$ for each run.}
    \label{fig:alpha_params}
\end{figure}

 During infall, the average disk-wide $\alpha$ grows on similar timescales to the RWI, with the initial growth scaling as $\tau_{\rm in}$. As the instability saturates, the magnitude of the disk-wide $\alpha$ approaches a constant value with time (Figure \ref{fig:alpha_params}). While more annular concentrated deposition develops robust $\alpha$ on slower timescales, this stream-like infall is on the whole just as effective at generating turbulence, likely in part due to the elevated turbulence generated within the infall zone for those cases. 

\begin{figure}
    \centering
    \includegraphics[width=0.9\columnwidth]{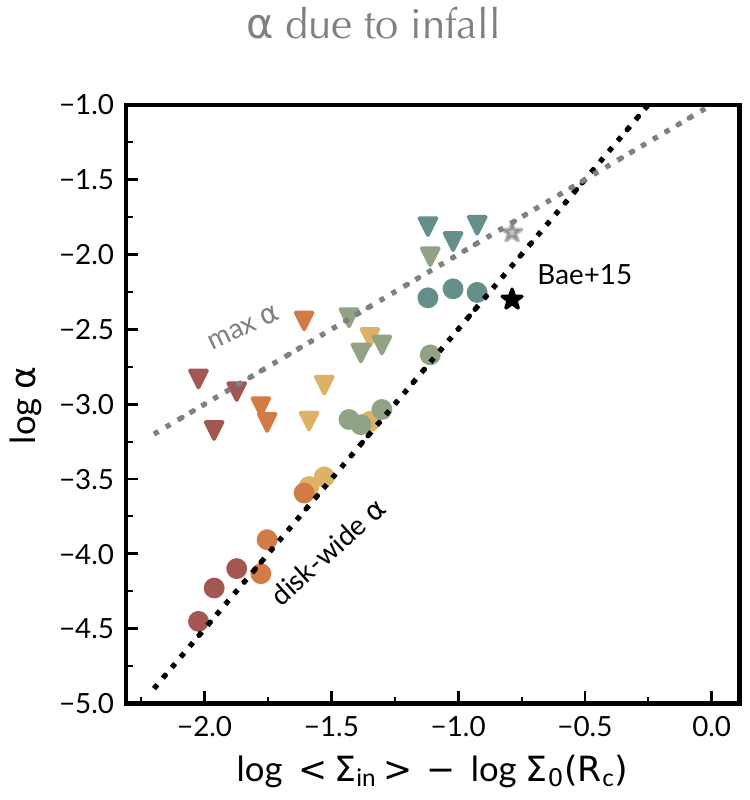}
    \caption{The disk-wide $\alpha$ (circles) and maximum
      $\alpha$ (triangles) at  $t = t_{\rm ep}$ for each
      model, plotted against the model's mass loading parameter $\log
      \langle\Sigma_{in}\rangle - \log \Sigma_0(\rc)$. Corresponding
      values from the standard run in  \citet{bae15} are plotted as
      starred markers. The black and gray dotted lines have slopes of
      2 and 1, respectively. Points are colored by their outer infall radius $\rc$ following Figure~\ref{fig:pspace}, \edit2{indicating that for models in this study the mass loading parameter increases with $\rc$}.}
    \label{fig:mass-loading}
\end{figure}

\par

    The strength of the RWI driven turbulence depends on the size of
    the initial perturbations produced by infall mass loading.
    Previous studies find the amount of angular momentum transport by
    spiral waves launched from vortices is directly correlated to
    their size \citep{vortex10}, which is influenced by both the scale
    height at vortex center and the relative amplitude of the initial
    surface density perturbation \citep{ono18}, both of which scale
    with the centrifugal radius of infall in this work. To measure the
    initial perturbations produced by infall, we
define a quantity for the relative mass loading

$\langle\Sigma_{\rm in}\rangle / \Sigma_0(\rc)$, the ratio of the average surface density perturbation per dynamical time
\begin{equation}
    \langle \Sigma_{\rm in}\rangle = \frac{\dot{M}_{\rm in}}{4\pi (\rc^2 - \rin^2)} \tau_{\rm c}
\end{equation}
to the initial disk surface density at the centrifugal radius
$\Sigma_0(\rc)$. Plotted in log-log space in Figure \ref{fig:mass-loading}, the disk-wide $\alpha$ at the end of infall has a tight relationship with the relative mass loading, where the black dotted reference line has the form:
\begin{equation}
    \log \alpha = 2 \ [ \log \langle\Sigma_{\rm in}\rangle - \log \Sigma_0(\rc)] - 0.5.
\end{equation}

Compared to the disk-wide $\alpha$ at the same simulation time, the
relationship to mass loading of the maximum $\alpha$ achieved in the disk has more scatter; the gray reference line for those points has a slope of unity, for comparison. We calculate the mass-loading ratio for the standard infall model in \citet{bae15} and find that their measured $\alpha \sim 10^{-3}$--$10^{-2}$, is also consistent with this scaling.

\par Overall, we find that in these models the magnitude of measured turbulence driven by infall is heavily dependent on whether it occurs primarily in the inner or outer disk, since the maximum achievable $\alpha$ is directly correlated with the outermost radius of the infall zone $\rc$. Outer disk infall drives the highest magnitude turbulence across the largest fraction of the disk area. 

The relationship in Figure \ref{fig:mass-loading} is 

      thus consistent with the scaling with $\rc$ expected from
      \citet{vortex10} and \citet{ono18}.

\subsection{Post-Infall Conditions}\label{sec:after_in}
\par
\begin{figure}[h!]
    \centering
    \includegraphics[width=0.9\columnwidth]{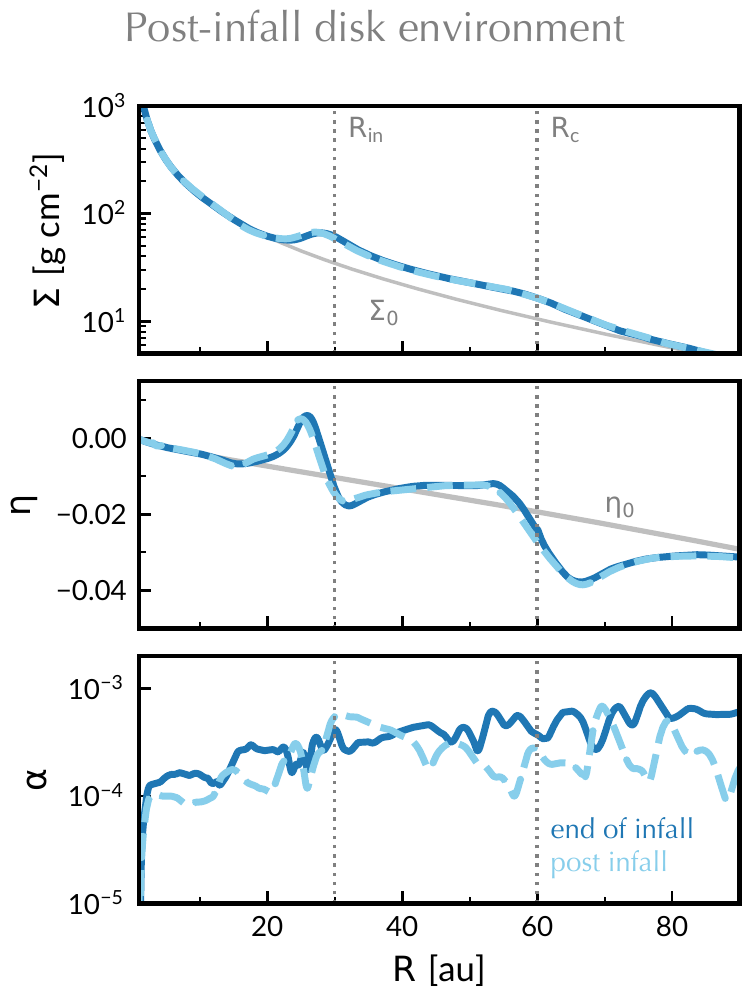}
    \caption{The surface density \textit{(top)}, pressure gradient parameterized as $\eta$ \textit{(center)}, and measured $\alpha$ parameter
\textit{(bottom)} for the fiducial run shown at the end of infall and post-infall, $t = 1.5\times 10^4 \unit{yrs}$ and $t = 3.5\times 10^4 \unit{yrs}$, in dark blue and light blue dashed lines, respectively. The initial profiles for $\Sigma_0$ and $\eta_0$ are shown in gray. }
    \label{fig:post-infall}
\end{figure}

 Each model was evolved for an additional $2 \times 10^4 \unit{yr}$
after $t_{\rm ep}$ in order to investigate the longevity of flow structures produced by infall.
While some structures show evidence of spreading, particularly for cases of outer disk infall, the radial profile of the surface density does not evolve significantly in the post-infall phase (see Figure \ref{fig:post-infall}). However, as shown in Figure \ref{fig:vortices}, the post-infall phase marks the start of vortex migration,  where individual vortices at $\rin$ or $\rc$ can be seen to have drifted inward from their initial positions. This basic migration behavior is consistent with other numerical simulations of Rossby vortices \citep{vortex10}, and has been attributed to a net exchange of angular momentum between the disk and vortex transported by spiral density waves due to a difference between the angular momentum flux at the inner and outer boundary of the vortex.

\par We estimate the inward migration rate of vortices by tracking the change in the radial location of the innermost peak in vorticity at $\rin$ and $\rc$, calculating the average migration speed over the post-infall period. As in previous studies, there is a correlation between migration rates and the scale height at vortex center $H_{\rm v}$. At $\rin$, inward vortex migration rates  
\begin{equation}
\begin{split}
\tau_{\rm v} & \sim 2 \times 10^{-8} \unit{au \ yr^{-1}} \left
  (\frac{H_{\rm v}}{0.05 \unit{au}}\right),\\
& \sim 2 \times 10^{-8}  \unit{au \ yr^{-1}} \left (\frac{\rin}{1
    \unit{au}}\right)^{5/2}
\end{split}
\end{equation}
have the same scaling with $H_{\rm v}$ to that of isolated vortices in \citet{vortex10}. 
 At $\rc$, however, we find that inward migration rates are generally slower than for inner infall zone vortices at comparable radii, but have a much steeper relationship with scale height,
\begin{equation}
    \begin{split}
        \tau_{\rm v} & \sim 3 \times 10^{-11} \unit{au \ yr^{-1}} \left (\frac{H_{\rm v}}{0.05 \unit{au}}\right)^3\\
        & \sim 3 \times 10^{-11} \unit{au \ yr^{-1}} \left (\frac{\rc}{1 \unit{au}}\right)^{15/4}.
    \end{split}
\end{equation}
 
\par During infall, the disk surface density is consistently perturbed, leading to the formation of multiple generations of RWI vortices, particularly for vortices at outer radii (Figure \ref{fig:vortices}).
In our case, angular momentum exchange is complicated by interactions between multiple generations of vortices, such that inward migration speeds of the first generation vortices are not direct proxies for angular momentum transport.  

\par In \citet{vortex10}, pressure bumps are argued
   to impede vortex migration, but our post-infall calculations are more consistent with \citet{ono18}, who find that pressure bumps induced self-consistently by the RWI do not prevent vortex migration. In this study, we find that RWI induced pressure perturbations tend to migrate along with their associated vortex. 

   However, the stalling effect of external pressure and surface density bumps on migration discussed in \citet{vortex10} could become important for the long-term evolution of vortices produced at $\rc$ and migrating through the infall zone. 

\par We do note that these comparisons to local box simulations are largely qualitative, as dedicated RWI studies induce the RWI with constant amplitude density perturbations, have constant numerical viscosity, and do not include the effects of additional velocity shear due to the impact of infalling flow streams; these additional effects reinforce the vortensity profile of the disk and locally alter the turbulent $\alpha$.

\par Based on the range of migration speeds \edit1{of the innermost vortex} $\sim 10^{-5}$--$10^{-3} \unit{au \ yr^{-1}}$ \edit1{varying with $R$ across a range of $\sim 10 - 100 \unit{au}$}, the fastest migration timescales would be on the order of several hundred kiloyears, with more typical estimates on the order of several megayears.

\par During infall, we can expect the turbulent $\alpha$, in part, to be locally driven by the momentum deposition of infalling material, particularly for stream-like infall. Post-infall, without additional turbulent forcing, we find that the average disk $\alpha$ is sustained at (or surpasses) its end of episode value for infall parameters $\rc > \rd$ (Figure \ref{fig:alpha_params}). 

The increase in $\alpha$ for outer disk infall is associated with infall models that host multiple generations of vortices, producing additional shear stresses.
\par However, $\alpha$ does decay post-infall for episodes of inner disk infall. The decay times for the value of the disk-wide $\alpha$ to drop by a factor of two, range from  $\tau_{\alpha} = 1$--$4 \times 10^4 \unit{yr}$. This decrease for inner disk infall models largely occurs at radii $R>\rc$ (Figure \ref{fig:post-infall}).

\subsection{Effects on Dust Evolution}\label{sec:trapping}

\par After infall, the dynamical landscape of the disk has been significantly altered. Following the model in \citet{birnstiel16}, we look at the relevant timescales for basic dust evolution. 
The speed of dust transport by turbulent diffusion is characterized by the diffusion timescale
\begin{equation}
    \tau_{\mathrm{diff}} = L^2 \frac{(1+\St^2)}{\nu},
  \end{equation}
where the characteristic length scale is $L$ and the viscosity \edit1{,$\nu$} is set by the total measured $\alpha$. Transport due to the inward radial drift of particles, induced by the pressure gradient (see Equation \ref{eq:eta}) is described by the radial drift timescale
\begin{equation}
    \tau_{\mathrm{drift}} = L \frac{(\St + \St^{-1})}{\eta v_{\rm k}}.
\end{equation}
We also compare
the collisional growth timescale due to particle sticking
\begin{equation}
    \tau_{\mathrm{growth}} = \frac{a \rho_{\rm s}}{\rho_{\rm d}\Delta v},
\end{equation}
where we assume relative collisional velocities due to turbulence

\begin{equation}\Delta v =  c_{\rm s} \left(\frac{3\alpha}{\St +
      \St^{-1}}\right)^{1/2}
\end{equation}
from \citet{oc07} and use the dust volume density with a vertical distribution set by settling
    \citep{birnstiel16}
\begin{equation}
    \rho_{\rm d} = \frac{\Sigma_g \epsilon}{H\sqrt{2\pi}}\left(\frac{\alpha}{\alpha + \St}\right)^{1/2}.
\end{equation}

\par
\begin{figure}[h!]
    \centering
    \includegraphics[width=\columnwidth]{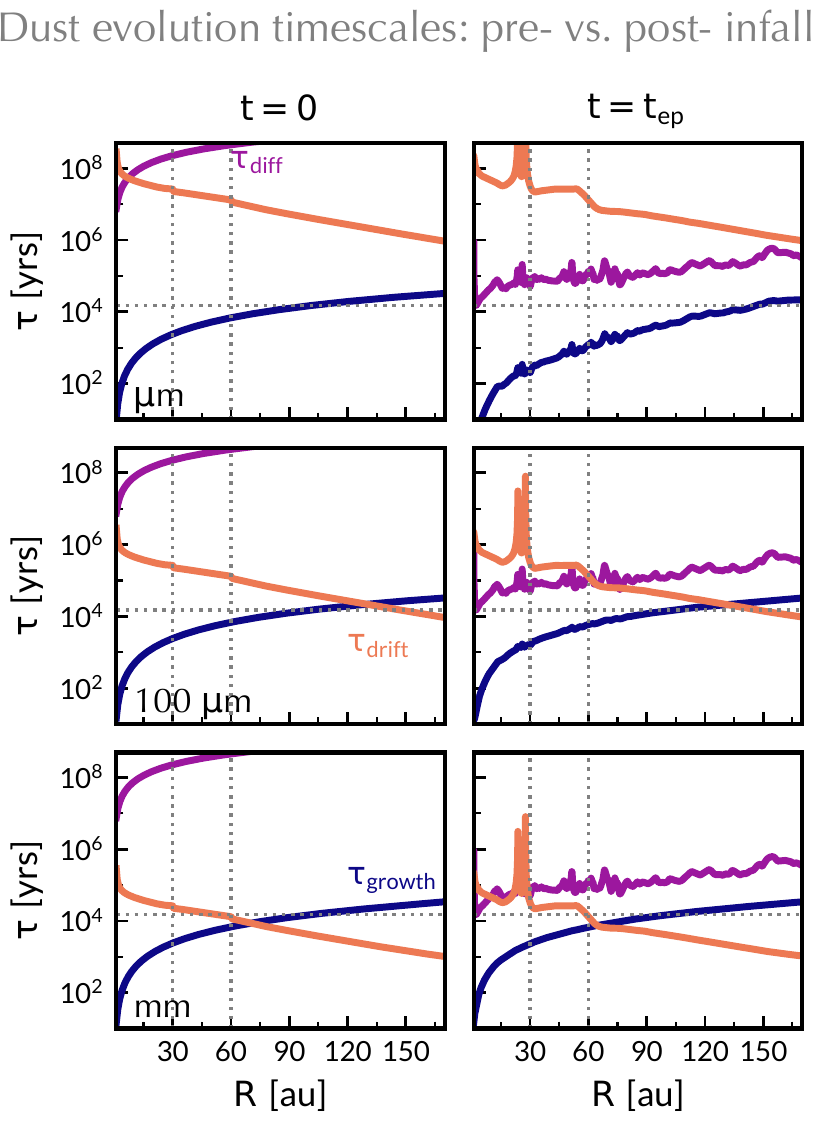}

    \caption{ Diffusion {\em (magenta)}, radial drift {\em (orange)},
           and collisional growth {\em (navy)} timescales at $t = 0$
           {\em (left)} and $t = t_{\rm ep}$ {\em (right)} for
           sheet-like infall with $\rc =\rd$ for 1~$\mu$m {\em (top)},
           100~$\mu$m {\em (middle)}, and 1~mm {\em (bottom)}
           dust. Small particles initially show long time-scales for
           drift and diffusion, but after the infall episode, disk perturbations and infall-induced turbulence shorten $\tau_{\rm diff}$, shorten $\tau_{\rm growth}$ in the outer disk for micron-sized particles, and lengthen $\tau_{\rm drift}$ within the pressure bump region. Dotted lines show {\em (vertical)} the locations of $\rin$ and $\rc$ and {\em (horizontal)} the characteristic dynamical timescale.}
     \label{fig:timescales}
\end{figure}

 In Figure \ref{fig:timescales}, we compare these timescales for
dust evolution at the start of the simulation and at $t= t_{\rm ep}$.  We
assume an initial dust-to-gas ratio $\epsilon = 0.01$ equal to the ISM value for three sizes of dust:
    1~$\mu$m, 100~$\mu$m, and 1~mm.
Using $L = 0.1 R$ as the operative length scale, we compare the times
over which one might expect changes in the radial dust distribution of at least $10\%$. 

 \par Comparing to the timescales over which we evolve the dynamics,
on the order of $\sim 10^4 \unit{yr}$, micron-sized grains would initially show limited growth at disk radii above $60 \unit{au}$. 
However, with infall, the growth timescale is shortened for small dust
out to nearly $120 \unit{au}$ for the fiducial run shown here. Larger
grains at outer radii have fairly rapid drift timescales, so
inward drift can push grains into growth dominated regimes. Without a
means to arrest radial drift, this leads to the standard picture of
drift limited growth. In the case of infall, however, rapid inward
drift from the outer disk could lead to pileup within pressure bumps,
potentially leading to accelerated growth there. The diffusion
timescales also become relatively more important within the inner disk
for small grains, although still on longer timescales than the
post-infall period we simulate here. While Figure~\ref{fig:timescales} shows approximate timescales, it suggests that the post-infall disk environment could be more conducive to planet formation than the typical steady-state disk.

    \par
Moving beyond estimates using the drift velocity, drift transport
depends on multiple components. In Equation \ref{eq:dustad},
perturbations to the dust surface density due to particle drift depend
on  (1) the local drift velocity (2) the differential drift velocity,
because of a traffic jam-like effect, and (3) the gradient in the surface density that dictates the bulk transport of material along the flow.  The main pressure bump corresponds to where $\ud > 0$, but most of the contribution to local drift arises from gradients in $\ud$, which peak near both $\rin$ and $\rc$, rather than just at the inward edge of the infall zone. 

\par Note that in this infall model, perturbations to the pressure
gradient occur not only due to perturbations in the gas surface
density, but also as a direct consequence
of the velocity shear induced by infalling material
(Eq. \ref{eq:j_in}).Thus, substructure in the gas surface density will not exactly match the amplitude and location of substructure in the dust surface density---a potential discriminating factor between rings induced by infall and those directly induced by planets, further discussed in \textsection~\ref{sec:discuss}.

\subsection{Rings and gaps}

\par We calculate the amplitude of relative perturbations in the dust density (\textsection \ref{sec:dust}) as the fractional change in dust surface density due to drift relative to an initially smooth surface density profile
\begin{equation}  \label{eq:A}
    A = \frac{\Sigma}{\Sigma_{0}} - 1
\end{equation}
 at $t=t_{ep} + 15 \unit{kyr}$. The results presented here are a
 conservative estimate of the \emph{relative} locations and amplitudes
 of potential dust substructures across our parameter space of
 simulations and possible dust sizes, without any model assumptions about
 the growth or fragmentation behavior of dust populations.

\begin{figure*}[ht!]
\centering
\gridline{\fig{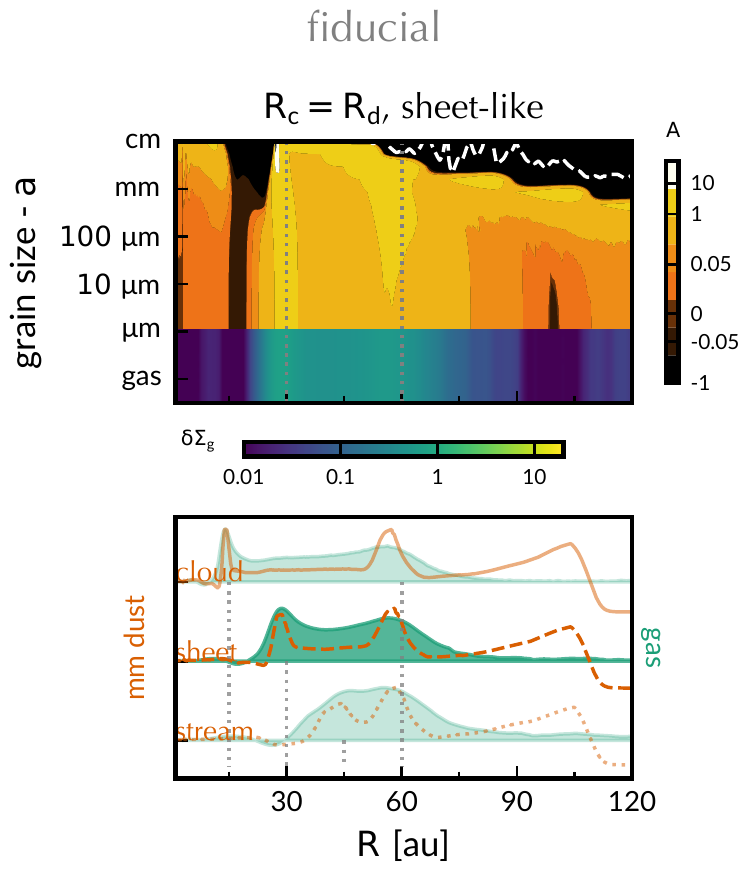}{0.4\textwidth}{}
 \fig{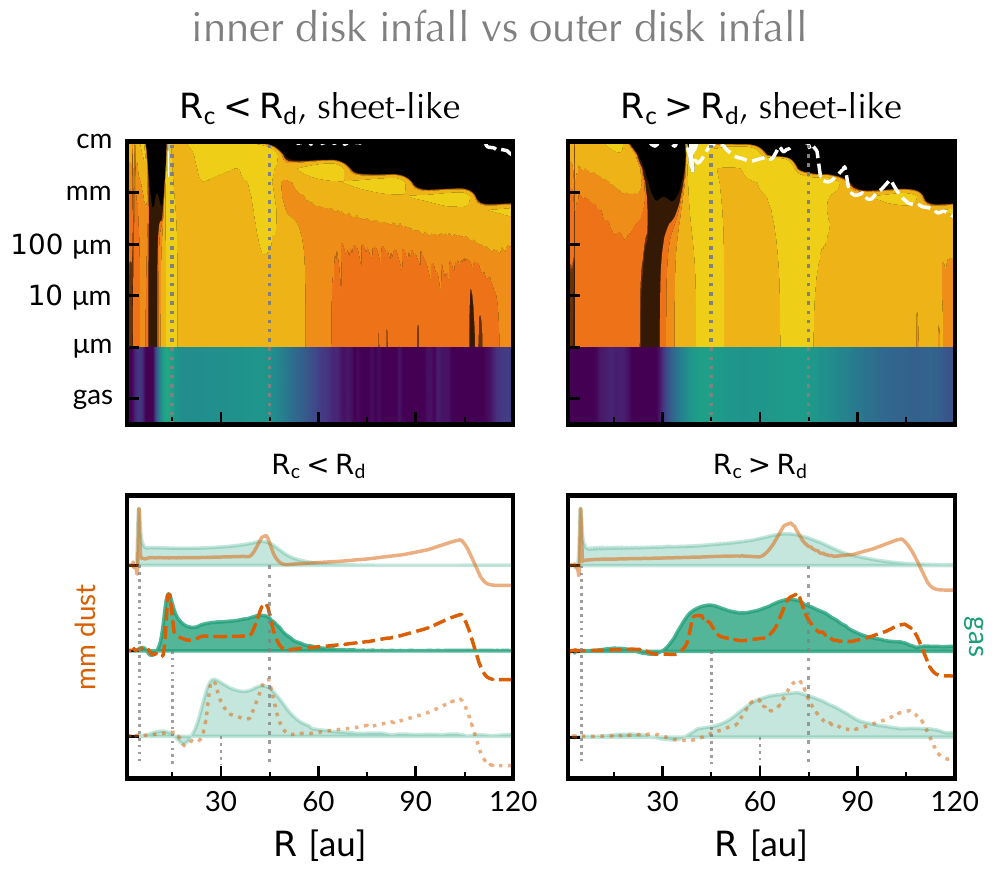}{0.53\textwidth}{}}
    \caption{ {\em (Top:)} Relative amplitude of dust surface-density perturbation $A$ (Eq.~\ref{eq:A})
as a result of local dust concentration in
pressure bumps for particle sizes in 16 bins. 
 {\em (Bottom:)} Relative
locations of rings of millimeter-sized dust {\em (orange lines)} compared to the gas surface density {\em (teal)},
normalized to the same amplitude. In each column, the middle plot {\em (dashed line)} shows the sheet-like infall of the top panel with the stream-like \edit1{{\em (dotted line)}} and cloud-like \edit1{{\em (solid line)}} models with the same $\rc$ shown for comparison. \edit1{Locations of $\rin$ for these models
are shown as dotted lines descending from their respective plots.} \textit{(Left:)} The fiducial run, sheet-like accretion with $\rc =\rd = 60 \unit{au}$. \textit{(Center:)} Inner disk, with $\rc = 45 \unit{au}$. \textit{(Right:)} outer disk, with $\rc = 75 \unit{au}$. The maximum grain size due to fragmentation $a_{\rm frag}$ is shown with a dashed white line. $\rin$ for each model and $\rc$  are denoted by gray dotted lines.
\label{fig:dust}
}
\end{figure*}

\par In Figure \ref{fig:dust}, we show the amplitude of perturbations
versus radius for selected runs post-infall to demonstrate the
relative locations of rings and gaps in the dust and gas for a range
of dust sizes. For reference, we plot the size corresponding to the fragmentation limit due to turbulent collisions \citep{birnstiel16}
\begin{equation}
    a_{\rm frag} \simeq 0.08 \frac{\Sigma_{\rm g}}{\rho_s \alpha}
    \left(\frac{v_{\rm frag}}{c_{\rm s}}\right)^2  
  \end{equation}
as a dashed white line, where $v_{\rm frag}$ is the fragmentation velocity, for which we use the threshold velocity $10 \unit{m~s^{-1}}$ for predominantly icy grains in the outer disk \citep{sticky15}. The turbulent fragmentation size limit extends down to millimeter sizes for the case of outer disk infall; the estimates of the perturbation amplitudes in those cases should be taken as an upper limit. The outer disk infall case, where $\alpha$ is constant outside of the infall zone (Figure \ref{fig:alpha_iz}),  also illustrates a physically self-consistent scenario in which the fragmentation barrier is tracing out the same region in $a$-$R$ space as the radial drift line. 

\par As expected, the largest dust sizes experience the most overall
inward drift. In most cases, this would lead to fairly compact dust
distributions, if not for the robust pressure bumps that serve as drift
barriers. Infall-induced pressure variations are strongest at $\rin$
compared to the pressure bumps at $\rc$, with the exception for cases
of stream-like infall. Large grains concentrate in rings near both
$\rin$ and $\rc$, migrating inward with time, such that most rings are
several astronomical units inward of their respective infall zone edges. In the dust profiles in the bottom panel of Figure \ref{fig:dust}, the outermost peak at $\sim 100 \unit{au}$ corresponds to the location of the dust line, demarcating the progression of radial drift for millimeter sized particles over the post-infall period. 

\par The width of dust ring structures is correlated with their
location in the disk. Consistent with dedicated studies of the RWI
\citep{ono18}, we find the width of structures correlates with the
disk scale height at peak center where the peak full width at half
maximum is roughly half of the scale height at its center. Relative amplitudes of the dust surface density perturbations have opposite trends depending on whether structures are in the outer or inner disk, where the trend for $A$ increases for $R < \rd$ and decreases for $R > \rd$. Relative to the gas, peaks in the dust are narrow and concentrated near the edges of the infall zone, whereas the gas surface density is dominated by a broad ring.

\section{Discussion} \label{sec:discuss}
\subsection{Hallmarks of Heterogeneous Infall}

\par The traditional analytic framework that informs our understanding
of disk formation and evolution \citep[e.g.][]{tsc84} does not produce
structured disks. Such infall of uniformly rotating spherical
envelopes produces smooth disks that grow outward with time such that
$\rc \propto t^3$. Previous studies of 1D dust evolution in such viscously evolving disks during the initial self-similar collapse stage find the smoothly varying infall of the traditional model to have little impact in the overall outcome of dust population growth \citep{birnstiel10}.

\par Numerical simulations that capture the onset of disk instabilities 
have established that the infall of envelope
material can have dynamical consequences relevant for disk transport
and planet formation processes \citep{lesur15,bae15}.
However, \citet{bae15} show that even with the ability to incite the RWI, monotonically increasing $\rc$ to mimic self-similar disk formation can still stall the instability when the migration rate outpaces the growth of the instability.

Without an understanding of how model assumptions fit into a larger disk
evolution framework it is difficult to contextualize what a typical
model system looks like and determine a range of probable effects. In
this work we focus on a parameter exploration of a self-consistent
model for the infall of star-forming material in order to constrain
the hallmarks of these model systems.

\par The growing evidence for heterogeneous accretion onto cores both in
observations \citep{tobin10, yen19, pineda20} and numerical simulations \citep{smith10,kuffmeier17,kuznetsova20} suggests that we should not merely view anisotropic infall as a potential mechanism for generating substructure, but rather acknowledge that infall is likely to be anisotropic in some fashion and that this will naturally lead to the formation of disk substructure, playing an important role in disk
evolution and planet formation.

\begin{figure*}[htb!]
    \centering
    \includegraphics[width=6.5in]{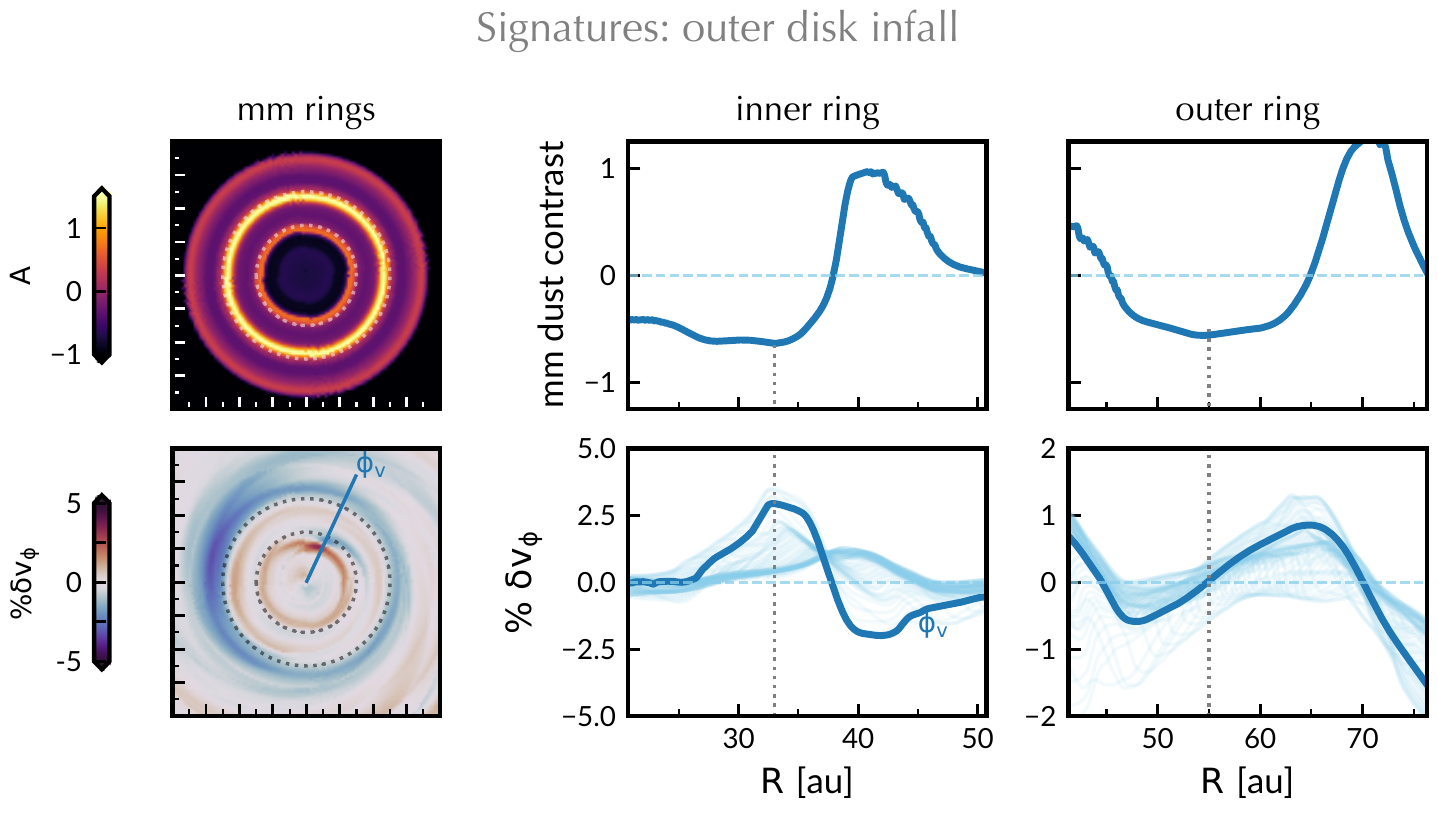}
    \caption{The relative location of millimeter-sized dust rings
     and kinematic deviations from the initial azimuthal gas velocity profile at a time of $1.5 \times 10^{4} \unit{yrs}$ post-infall,
with the original locations of $\rin$ and $\rc$ denoted by dotted
annuli {\it (left)}, are such that the minimum of the dust gap, the
location of which is marked in both panels by a vertical gray dotted
line, near the inner ring  corresponds with the peak in
super-Keplerian velocity as expected for localized vortex structures
{\it (center)}, but the apparent gap in front of the outer ring is
centered where deviations are zero, with the pattern more akin to that
produced by spiral wakes associated with a planetary companion {\it
  (right)}. Bottom center and right panels show sample radial profiles
of the velocity deviation (light blue), with the navy bold curve
corresponding to the profile at the vortex azimuth $\phi_{\rm v}$}
    \label{fig:sigs}
\end{figure*}

\par As a caveat, the range of parameters used here is based on the results from numerical simulations that, due to limitations in resolution, are biased toward higher mass cores; thus, our choice of initial star and disk model is kept constant, but consistent with the kinds of systems whose substructures have been well-characterized in the radio \citep[e.g.][]{dsharp18, sheehan20} which also tend to be higher mass and relatively bright systems. 

\par From our study, we have highlighted some of the main features of
a system that is undergoing or has undergone an infall episode: a broad gaseous bump across the infall zone with a prominent drift barrier ($\eta > 0$) at $\rin$ due to the presence of an inner vortex and a weaker barrier due to a vortex at $\rc$. Such vortices launch spiral waves, which can entrain dust in narrower structures \citep{bae15}, and have been shown to produce contrasts similar to spiral waves driven by low mass planets (up to tens of Earth masses) \citep{huang19}. 

\par Post-infall, differential radial drift contributes to the pile-up of mm-size dust at the edges of the infall zone. For infall in the outer disk, this results in the formation of rings outside of the nominal disk radius, $\rd$. 
In Figure \ref{fig:sigs}, we show an example comparison of the locations of millimeter-sized dust rings and gaps
with velocity deviations in the gas.
  This comparison demonstrates that vortices produce kinematic
deviations like those expected for planets \citep{dd20}, where spiral wakes produce strong deviations on the order of $5\%$.
Vortices rotate opposite to the flows perturbed by planetary companions and thus their kinematic signatures are expected to be easily differentiated from planetary perturbers. While this is true for the vortices near the inner ring, the velocity deviations near the outer ring at $\rc$ are centered on the gap between the two rings, potentially producing the appearance of deviations akin to those from a planetary perturber. 

\par The predicted appearance of dust substructures due to the
presence of a gap-opening planet has been well-described
\citep{dongfung17}, and analysis of the substructures in later phase
sources from the DSHARP survey have yielded an inferred population of
embedded planets \citep{zhang18} with planet mass, $\alpha$, and $H/R$
used as free parameters. However, without accurate tracers of both the
dust and the gas, the presence of continuum rings alone is not unique
to the disk-planet interaction scenario. For instance, studies of the
molecular gas of the substructured source GM Auriga have found the
potential for the gravitational instability \citep{gmaur21a} and
complex structure in the surrounding gas pointing to late-stage infall
\citep{gmaur21b}.

\par The one-dimensional dust evolution calculation presented here neglects non-axisymmetry, but RWI vortices have been shown to be effective for trapping grains of at least sub-mm sizes in non-axisymmetric structures \citep{meheut12,barge17}. Rings and non-axisymmetric vortex structures are likely to be most prominent for outer disk infall, given that the radial extent of vortices is dependent on the scale height at their location. This trend is consistent with the prevalence of non-axisymmetric features observed largely on the outskirts of disks \citep{marel20}. Planetary companions have been invoked to explain the presence of gaps and rings, however generating the large scale structures seen in \citet{marel20} with a planetary perturber requires sub-stellar mass companions ($M > 15 M_{\rm J}$).

\subsection{Anisotropic infall makes small disks?}

\par One could expect that as dust ring architectures are relatively common for the annular infall modeled in this study, they should be prevalent among embedded sources. However, the VANDAM survey of Class 0/I sources in Orion is dominated by compact unstructured sources \citep{tobin2020}; whereas substructures in the sample appear exclusively as large-scale exterior rings \citep{sheehan20}. The typical effective dust radii of these systems is less than $100 \unit{au}$, with most single sources having inferred disk sizes less than even $50 \unit{au}$, smaller than that expected from a rotating collapse scenario. 

\begin{figure*}[ht!]
\centering
\gridline{\fig{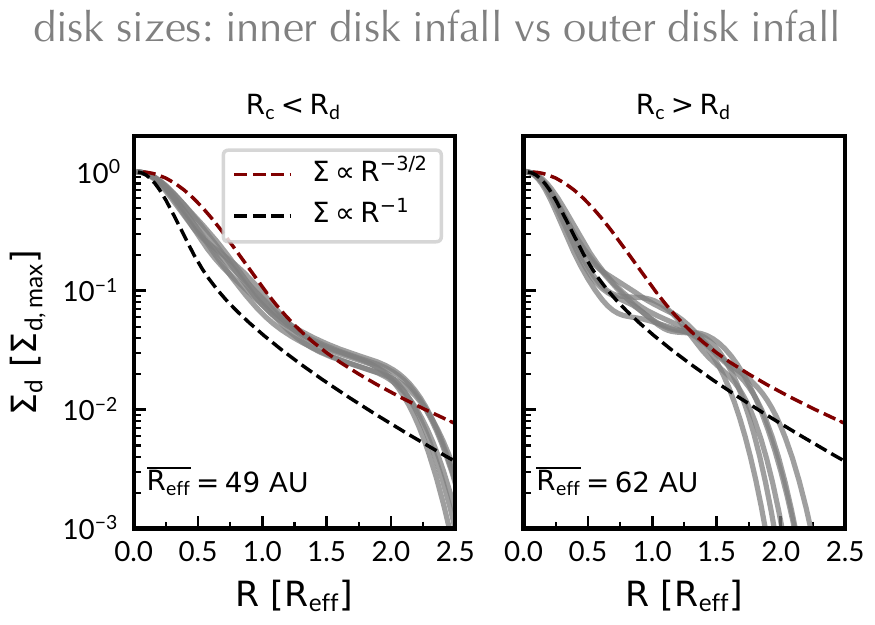}{0.45\textwidth}{}
 \fig{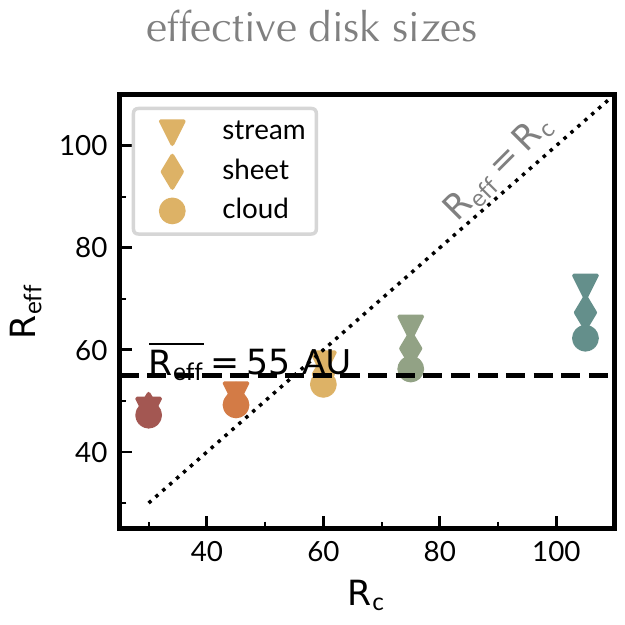}{0.3\textwidth}{}}
    \caption{\edit1{Dust} surface density profiles (gray) for inner vs outer disk infall, smoothed over $10 \unit{au}$ {\it (left)}. The effective disk size $R_{\rm eff}$ denotes where the mass enclosed is $68 \%$ of the total mass. Dashed lines denote surface density power-law profiles, where outer disk infall is best fit by the initial $-1$ profile (black dashed line) and inner disk infall is more consistent with a steeper $-3/2$ power-law (maroon dashed line). The mean effective disk size is 55 au, systematically smaller than $\rd$, with a weak dependence on the $\rc$ of the infall zone {\it (right)}. }
    \label{fig:sizes}
\end{figure*}

\par Looking at the evolution of millimeter grain surface density due to radial drift (Figure \ref{fig:dust}), both outer and inner disk infall show similar relative enhancement of the dust density at the pressure bumps. However, outer disk infall, which occurs outside of the disk critical radius $\rd$, deposits material in an initially less dense part of the disk over a larger surface area. A relative enhancement of a factor of two is still a fairly low absolute
surface density when it occurs in the outer disk, but a more significant concentration of mass in the inner disk.

\par In Figure \ref{fig:sizes}, we compare dust disk surface density profiles, smoothed over $10 \unit{au}$, for inner and outer disk infall. As an analogue to radii inferred from continuum observations, we denote the location of the effective radius $R_{\rm eff}$ as where the enclosed mass of the smoothed profile contain $68 \%$ of the total mass, such that the  unperturbed initial surface density profile has $R_{\rm eff} = \rd$. Inner disk infall produces systematically smaller disks as dust drift creates pile-ups near $\rc$. In addition, surface density profiles are more consistent with a steeper power-law $\Sigma \propto R^{-3/2}$. Outer disk infall is less likely to affect the overall fit of the surface density profile, tracing out the original disk radius $\rd$, as perturbations to the surface density profile then occur well outside of $R_{\rm eff}$. Anisotropic infall, where inner and outer disk infall is just as likely, accounting for radial drift, produces mean dust disk sizes $R_{\rm eff} \sim 55 \unit{au}$; consistent with an observed population of protostellar sources dominated by compact dust disks. Substructured disks in the VANDAM survey tend to be larger than the mean disk sizes in the sample, given that their radii are influenced by the location of the outer dust ring. As outer disk infall closer to $\rd$ is less likely to be resolved as a ring-like structure, large scale exterior substructures like those in \citep{sheehan20} should in fact be rarer among the population of observed early phase objects.

\subsection{Prospects for Disk Evolution}
\par This study evolves the disk post-infall for several tens of kiloyears, relatively short compared to the megayear timescales over which we expect protoplanetary disks to evolve. The results of this study provide a starting point for further investigation into the long-term consequences of anistropic infall on disk evolution. 
\par Notably, our results show a variety of radial profiles of the $\alpha$ parameter across the disk, increasing outside the infall zone for outer disk infall and decreasing for inner disk infall (Figure \ref{fig:alpha_iz}). In the steady-state $\alpha$-disk model
, which assumes a constant disk-wide $\alpha$ to parameterize viscosity,
    the viscous timescale $t_\nu \sim R^2/(\alpha c_{\rm s} H)$
    depends on radius as $t_\nu \sim R$ for a disk with a temperature structure set by stellar heating $T\propto R^{1/2}$. Given the radial dependence in $\alpha$
       that we find
for outer disk infall, one could expect the same viscous timescale at all radii, compared to inner disk infall which would only have comparable transport times interior to the infall zone. 
\par \citet{bae14} has shown that isotropic infall could power accretion outbursts on short timescales during disk formation. Accelerated transport across the whole disk due to anisotropic infall in the outer disk has consequences for the protostellar accretion and outburst behavior of early phase sources. Characterizing transport and outburst behavior due to anisotropic infall will require treatment of heating and cooling and the inclusion of disk self-gravity to capture fragmentation.
\par Within the parameter space and initial disk parameters explored here, we do not expect disk self-gravity to play an important role. At the end of infall, the Toomre Q parameter is well above the marginal stability criterion. In our study, by the time Q reaches a value at which self-gravity could affect the stability of high mode number vortices $Q < R/H$ \citep{lovelace13}, the instability has already saturated. However, the chosen $\dot{M}_{\rm in}$ is marginal, and higher infall rates will be in the regime at which self-gravity will need to be considered for both the RWI and gravitational instability.

\citet{bae15} show that quenching of the infall-induced RWI and dissipation of vortex structures can occur in the self-gravitating regime. Population level predictions for a full range of evolutionary outcomes, such as those originally put forth in \citet{kratter08}, including the effects of anisotropic infall will require including disk self-gravity and expanding the parameter space. 
\par Even as anisotropic infall creates regions of enhanced angular momentum transport, the results demonstrate that it also sets up barriers to dust transport by drift, robustly trapping most particle sizes at $\rin$. Over longer timescales, the pressure bump at $\rin$ could effectively segregate dust populations in the disk. Multiple episodes of infall from different accretion reservoirs may lead to compositional heterogeneity within the disk, a proposed mechanism for accounting for the isotopic variations in solar system chondrites \citep{piani21}. Investigating how anisotropic infall may affect the long-term compositional evolution of protoplanet bodies will require more sophisticated modeling of the dust evolution, accounting for growth and fragmentation over longer timescales. 

\section{Summary} \label{sec:conclude}
\par Growing evidence from both observations and numerical simulations suggests that disks are fed by anisotropic streams of infalling material, counter to the self-similar theoretical approach for disk formation and evolution. We performed a parameter study of anistropic infall onto disks, with model parameters informed by the properties of protostellar core accretion events from large-scale star cluster formation simulations \citep{kuznetsova20}. The parameter space subtends  accretion flow geometry from the nearly spherical to the stream-like and landing zones within the inner disk and in the outer disk. Results show that anisotropic infall significantly perturbs the disk, generating perturbations strong enough to incite the Rossby wave instability within a few dynamical times at the outermost radius of infall. 
\par Infalling material generates shear stresses, with the highest measured viscosity parameter $\alpha \sim 10^{-2}$ occurring in cases of outer disk infall. How locally generated mechanical turbulent stresses affect long-term disk transport and evolution and potentially induce outbursts is a subject of further study.

\par Infall-induced azimuthal shear is robust near the edges of the infall zone, allowing entrapment of the largest dust particles, with the potential for accelerated planet formation at early times. We post-process simulations, calculating the one-dimensional dust evolution due to radial drift. We find that ring formation due to differential drift can occur rapidly in the post-infall phase, with extended substructures likely for infall occurring in the outer disk. Infall in the inner disk, equally likely in our theoretical framework, accumulates dust in the infall zone, resulting in a compact dust disk, consistent with measurements of small sub-millimeter dust radii in embedded sources. 
\par The anisotropic infall model presented here results in the ready formation and reinforcement of pressure bumps in the disk during the earlier phases of protostellar disk evolution. The presence of multiple drift barriers likely has consequences on the long term evolution of the disk dynamics and on the bulk composition and sizes of planetary bodies segregated by drift barriers. 

\acknowledgments
Support for A.K. was provided by NASA through the NASA Hubble Fellowship grant \#HST-HF2-51463.001-A awarded by the Space Telescope Science Institute, which is operated by the Association of Universities for Research in Astronomy, Incorporated, under NASA contract NAS5-26555. M.-M.M.L. was partially supported by NSF grant AST18-15461. This work used the HPC \emph{Mendel} cluster at the American Museum of Natural History developed with National Science Foundation (NSF) Campus Cyberinfrastructure support through Award\#1925590 and the Extreme Science and Engineering Discovery Environment (XSEDE) \emph{Bridges} at the Pittsburgh Supercomputing Center under allocation AST190025, which is supported by NSF grant number ACI-1548562.



\appendix
\section{Effect of Numerical Resolution}

\par
\begin{figure}[h!]
    \centering
    \includegraphics{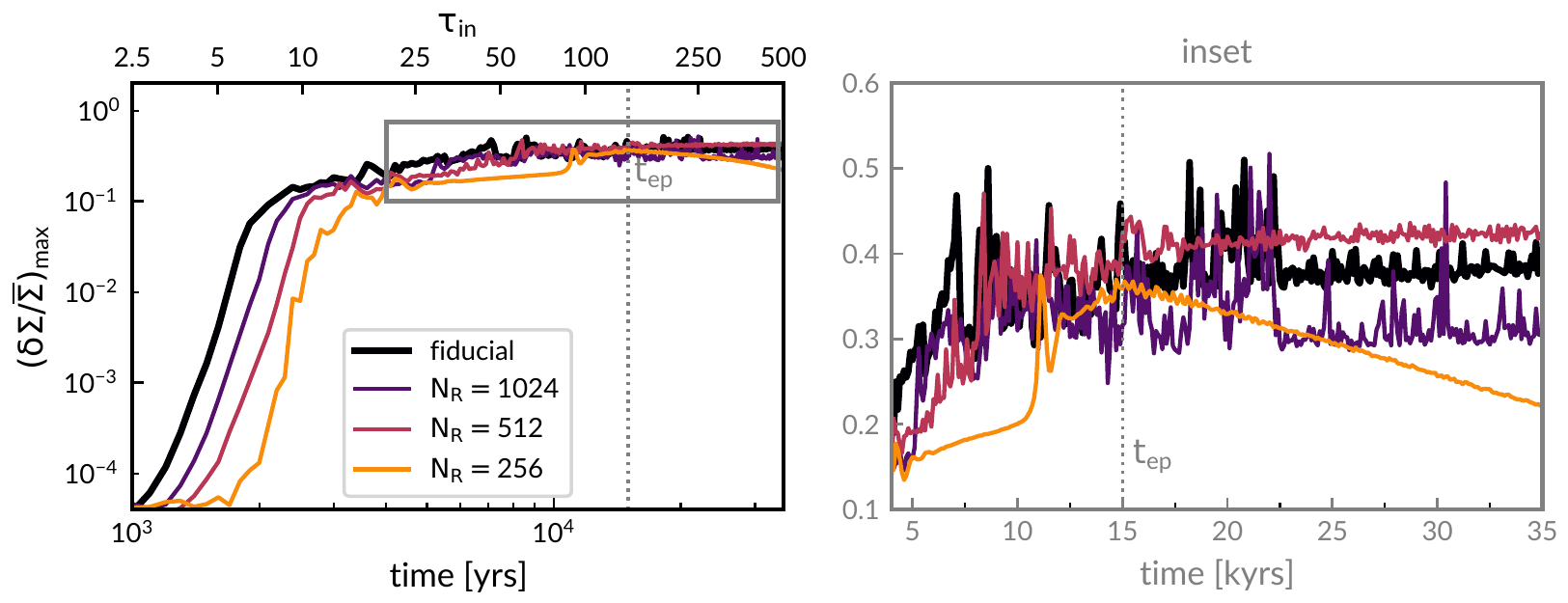}
    \caption{The evolution of the maximum non-axisymmetric surface density perturbation over the simulation time, as in Figure \ref{fig:t_rwi}, for the numerical resolutions tested, with the fiducial run ($N_{R} = 2048$) shown as a bold black line. The launch time of the instability is delayed for lower resolutions. The inset region marked by the gray box in the left panel is magnified on the right and 
    shown on a linear scale. 
    }
    \label{fig:app_perturb}
\end{figure}

 The simulations in this study were performed with a resolution of $N_{\rm R} = 2048, N_{\phi} = 1024$, on a logarithmic radial grid corresponding to a constant $\Delta R/R \sim 0.0027$. Here, we compare radially logarithmic grids of varying resolution, $N_{\rm R} = 256, 512, 1024, 2048$ with $N_{\phi}= 128, 256, 512, 1024$, respectively, for the fiducial infall model used above: $\rin = 30 \unit{au}, \rc = 60 \unit{au}$. 
When comparing the development of the RWI over time using the maximum relative non-axisymmetric perturbation strength, as in Figure \ref{fig:t_rwi}, we find that the resolution affects the launch time of the instability
, as well as the perturbation strength during the non-linear phase (Figure \ref{fig:app_perturb}).

\par
\begin{figure}[h!]
    \centering
    \includegraphics{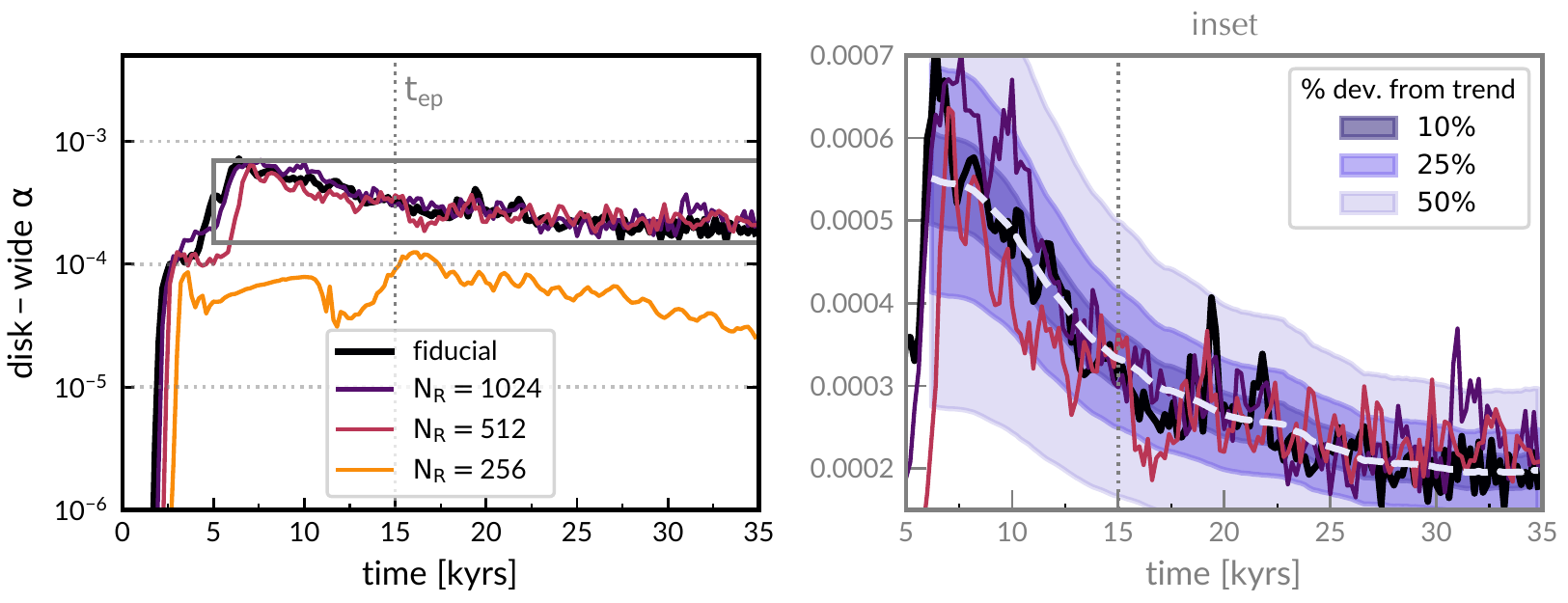}
    \caption{The time evolution of the disk-wide $\alpha$ parameter at different numerical resolutions, with the fiducial run ($N_{R} = 2048$) shown as a bold black line. The inset region marked by the gray box in the left panel is magnified on the right and 
    shown on a linear scale. ({\em inset}) A smoothed trend line corresponding to  the fiducial run is overplotted as a dashed 
      white
    line, with contours for $10\%, 25 \%,$ and $50\%$ deviation from the trend shown in shades of violet.}
    \label{fig:app_alpha}
\end{figure}

 Notably, for the lowest resolution shown for comparison $N_{\rm R} = 512, N_{\phi}= 256$, the perturbation strength decays after infall ends. At that resolution, the RWI at $\rc$ is quenched, producing no long-lived vortices. This effect can be seen in the evolution of the viscosity parameter over time for the lowest resolution case, wherein the disk-wide $\alpha$ does not see the same secondary growth period after $4 \unit{kyr}$ corresponding to the growth of the RWI at $\rc$. Compared to the overall trend of the decay of the disk-wide $\alpha$ of the fiducial run, shown as a smoothed trend-line in Figure \ref{fig:app_alpha}, the deviations introduced by the time variation of the fiducial disk-wide $\alpha$ is usually on the level of 10--25\%. The magnitude of errors introduced at lower resolutions can be twice that of the noise level in the fiducial case. 

\par
\begin{figure}[h!]
    \centering
    \includegraphics{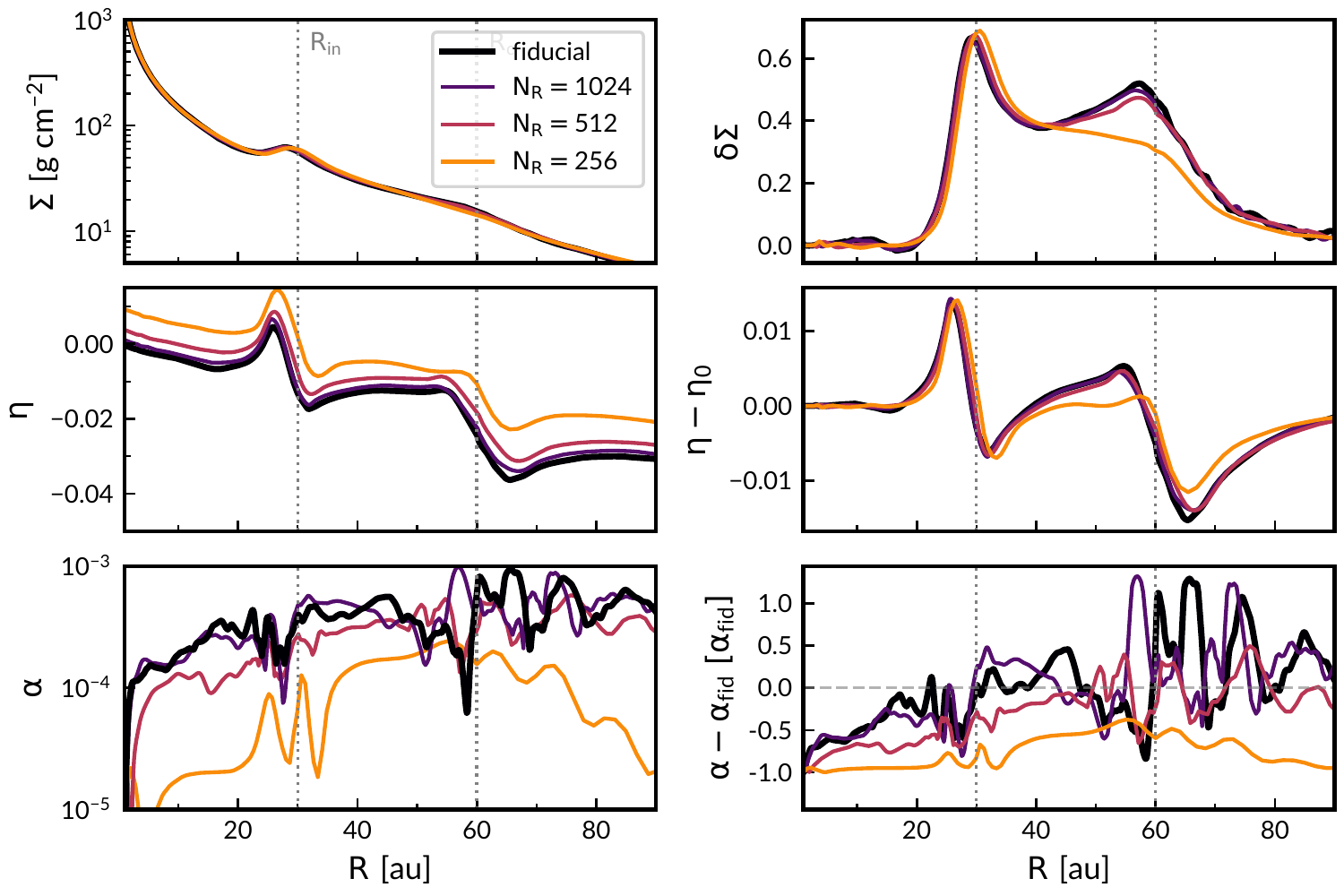}
    \caption{ ({\em left}) Metrics of disk dynamics shown in Figure \ref{fig:post-infall} across different numerical resolutions selected during the infall period, $t < t_{\rm ep}$, from top to bottom: surface density, pressure gradient parameter $\eta$, and measured $\alpha$ parameters. ({\em right}) Deviations in same metrics, from top to bottom: the perturbation to the surface density $\delta \Sigma = \Sigma/ \Sigma_0 - 1$, perturbation to the pressure gradient $\eta - \eta_0$, and perturbation relative to the disk-wide $\alpha$ parameter measured for the fiducial run at the selected time, $\alpha_{\rm fid} = 3.86 \times 10^{-4}$.}
    \label{fig:app_dynamics}
\end{figure}

 Across the metrics introduced in Figure \ref{fig:post-infall}, we show 
   in Figure~\ref{fig:app_dynamics} that quenching of the RWI by low numerical resolution
at the outer radius results in a relatively lower surface density (and pressure) bump near $\rc$. At higher resolutions, the relative perturbation in the pressure gradient compared to its initial value $\eta - \eta_0$ is not strongly affected by resolution. 

However, the value of the large scale pressure gradient represented by $\eta$ does depend on how well the pressure gradient is resolved. Though $\eta$ sets the value of the dust drift velocity $u_{\rm d}$ and associated drift timescales, the formation of dust gaps and rings depends much more on the differential drift than it does the absolute value of the drift velocity. The primary effect of under-resolving the pressure gradient is realized at smaller radii, where the direction of the pressure gradient is reversed, resulting in artificially super-Keplerian flow in the inner disk where $\eta > 0$. 

\newpage

\bibliography{refs}{}
\bibliographystyle{aasjournal}



\end{document}